# Stratigraphic forward modeling software package for research and education

by Daniel M. Tetzlaff[1]

## Abstract

We have developed an open-source software package ("SedSimple") for stratigraphic forward modeling. It is aimed for use in research and education. The package uses ASCII files for input and output to facilitate understanding, debugging and preparation of multiple data sets though simple user-developed scripts. A free graphic program is available to assist in editing input and rendering results in 3D plus geologic time animation. The open-source concept offers the ability to understand the algorithm in detail, and modify or extend it as needed.

Several algorithms are available in the program to model various sedimentary environments and processes, including alluvial, fluvial, turbiditic, carbonatic, deep-sea clastic, wave action, vertical tectonics and sea-level changes. These processes can be run concurrently. Furthermore, the effect of some processes can be made to affect others (for example wave action can affect carbonate growth, while also causing erosion and redistribution of the deposits).

The algorithms are fully deterministic. As most forward models of natural environments, the results cannot be readily conditioned to data. However, the model is much faster than similar commercial packages of the same resolution. Therefore, we foresee that one of the main uses of the package will be to study workflows that require multiple realizations to obtain statistical predictions, such as Monte Carlo methods, combination with geostatistical methods, artificial intelligence and machine learning that require multiple inputs of geologically realistic models.

---

1   Westchase Software Corporation, Houston, Texas, U.S.A., website: https://www.wsoftc.com, email: dan@wsoftc.com

# Introduction

"Stratigraphic Forward Modeling" (SFM), "Quantitative Dynamic Stratigraphy", and "Geologic Process Modeling" are just a few of the names used to describe a rapidly growing set of methods used in the oil industry and in research to model the erosion, transport, deposition, diagenesis and deformation of sedimentary deposits. Their goal is to understand stratigraphic architecture through genetic processes rather than through interpolation based on spatial statistics. The processes that must be modeled to obtain useful results are complex and researchers often need to implement their own algorithms, a requirement for which commercial software tends to be ill prepared. Researchers and students often need to modify existing algorithms or create new ones to implement a concept not contained in existing software. The open-source concept fulfills this need, in addition to making the software available at no cost.

Simulation, as employed here, is the process of creating a mathematical model of an actual system and performing experiments with the model that imitate the actual system's behavior. Employing this meaning, simulation involves a broad set of techniques, applied in every field of science, and treated by many authors. Examples of simulation in the earth sciences include representation of flow in groundwater reservoirs and in oil and gas reservoirs, flood routing and river-stage calculations, and sedimentation modeling.

## Classification

Geologic simulation models can be classified with respect to whether they are experimental or theoretical, probabilistic or deterministic, non-dimensional, one-, two-, or three-dimensional, static or dynamic. SedSimple is a dynamic deterministic model.

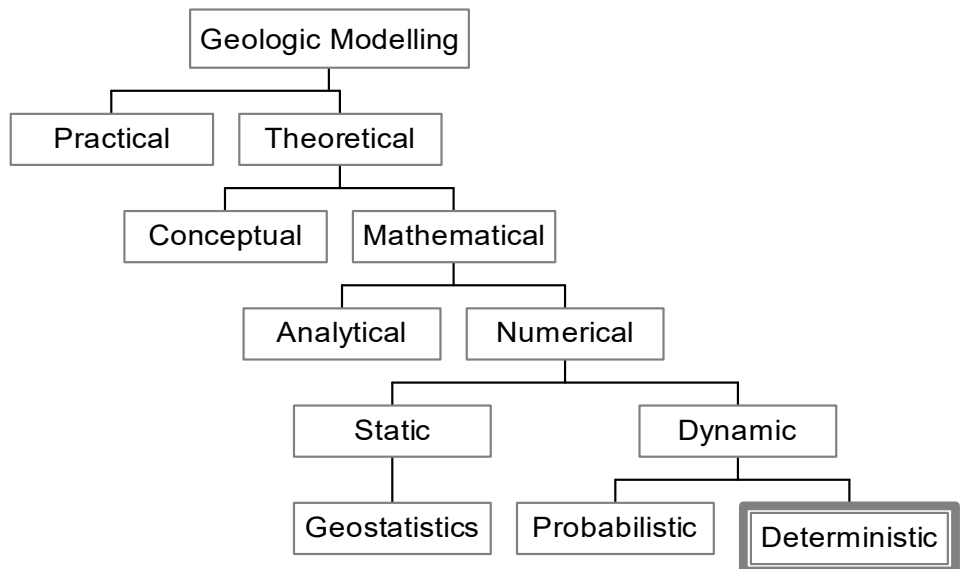

*Figure 1: Schematic classification of geoogic process models.*

Broadly, simulation models can be divided into two groups, namely (1) physical models, that embody small-scale versions of actual physical processes, such as flumes or sand-filled tanks in a sedimentation laboratory, and (2) mathematical models, in which relationships are represented as equations and logic statements, generally as computer programs.

Physical models have an advantage in that they directly represent the same phenomena as the actual system, but are often hindered because some of the physical variables represented cannot be scaled downward or upward. Mathematical models, on the other hand, do not suffer from scaling limitations, but their fidelity may be limited by the simplification necessary for treatment by computer, particularly the subdivision of a continuous medium into a finite number of cells or steps.

Simulation models may contain elements of chance, as when cause-and-effect relationships are influenced or governed by probability distributions, (for example, sedimentary sequences represented by Markov chains), or they can be totally deterministic. However, even if a simulation model is entirely deterministic, its behavior may be "pseudorandom" or virtually probabilistic, in that a small change in initial conditions may result in significant change in the model's behavior. Thus the model may seem to behave in random fashion even though it contains no random elements. For example, some models can simulate the development of a braided stream. Beginning with a straight channel, minor changes in flow may later cause a significant deviation in the channel. If initial conditions are changed slightly, the later form of the channel may be very different. Such a model is pseudorandom in the sense that even though it contains no elements designed to cause it to behave in a random fashion, its behavior can only be predicted if the initial state is known in perfect detail.

If time is represented in a simulation model, it can be called a dynamic model or process model. Dynamic models are important in earth sciences because they may provide an insight into the past events that produced features observed at present. For example, the nature and magnitude of processes that filled an evaporite basin can be analyzed with dynamic models. Simulation experiments can be performed under various hypothesized flow circulation patterns in the basin, and with different sea water inflow conditions, until the actual stratigraphic configuration can be reproduced by the model. Truly dynamic models permit the realization of experiments that can not only shed light on the processes that generated a clastic sequence, but also predict the configuration of sedimentary deposits between or beyond areas for which information is available, as for example, in an oil field where wells provide only local information.

## Uses

Simulation of sedimentation is difficult because of the complexity of the systems represented. Although a model can be much simpler than the real system, the model should represent all basic relationships between the system's components of specific interest. Such goal, however, may be difficult to attain in complex systems. Analytical solutions for systems of equations that govern flow and clastic sedimentation may not be feasible. Even numerical methods consume a large amount of computer time for relatively simple models. Some models of processes that transport clastic sediments work satisfactorily, as described below, but in general they are restricted to particular sedimentary environments or conditions, or they simulate only short periods of time.

We need process models that can simulate clastic processes in large sedimentary systems over extended periods of time. Such models could be used in at least three major applications:

1. To better understand mechanisms that create depositional sedimentary features, particularly the reciprocal interaction between flow and topographic configuration.

2. To reconstruct the depositional history of a sedimentary sequence and to predict its configuration beyond the locations where it can be observed, as in wells or outcrops. For example, there may be several alternative hypotheses concerning ancient geographic conditions that influenced the deposition of the sedimentary sequence observed in a well. A process model can be used to simulate the consequences that stem from each of the alternative hypotheses. Flow rates, sediment input, and basin slope can be adjusted until the simulated deposits and observed deposits are in suitable accord. If a feasible model can be established, it can be considered for use in predicting the sedimentary deposits in locations where geological or geophysical information is lacking.

3. To predict the short term future behavior of fluvial systems, as for example in dealing with engineering problems posed by erosion and deposition along river banks or man-made channels, and sedimentation in reservoirs. Although procedures exist that adequately handle some of these problems, few procedures are useful in cases where channel shifts and other major topographic changes take place.

The challenge in simulating sedimentation processes lies chiefly in the fact that medium or large scale sedimentary features are not the result of isolated processes, but instead have been produced by the interactions between flow, sediment transport, and topography resulting from erosion and deposition.

While we have a good understanding of how flow causes erosion, transport, and deposition, and conversely how topographic features affect flow, the mutual interaction of these processes is known mostly conceptually. SedSimple focuses on the role of sedimentary processes within a coherent system, which can be monitored quantitatively.

## History

Stratigraphic forward modeling did not start as a method for modeling the present day configuration of stratigraphic sequences, but as a means to understand how sedimentary processes occur. Early efforts to quantify sedimentary processes go back to to geomorphologists. Around the mid $20^{th}$ century quantitative models were developed to simulate the evolution of a graded river's profile. They were based on local relationships of slope, sediment load and erosion vs. deposition to predict the overall evolution of a river. However, these models were essentially one-dimensional and did not result in stratigraphic predictions.

Computer simulation models have also been devised to deal with engineering applications of processes that involve scour and deposition in man-made channels and reservoirs. One notable early such model was developed by the Hydrologic Engineering Center (HEC) of the U.S. Army Corps of Engineers. The computer program, called HEC-6, calculates flow conditions in a river or channel, and predicts erosion, transport and deposition of clastic sediment. Program input consists of information on channel geometry, water discharge, and sediment characteristics. Channel geometry is specified by a series of vertical transverse cross sections. The horizontal location of the stream's bed is assumed to be fixed, but the program allows to model a river system with tributaries and variable water discharge and sediment input.

Probably the first model developed specifically for stratigraphic configuration and its response to sea-level change and sediment input was developed by Harbaugh and Bonham-Carter (1981), to represent sedimentation on a continental shelf. These models were based on an earlier conceptual model. Their model assumes that the continental shelf can be represented as a series of columns that extend from the shore toward a basin. The model is thus two dimensional, and is represented only in cross section:

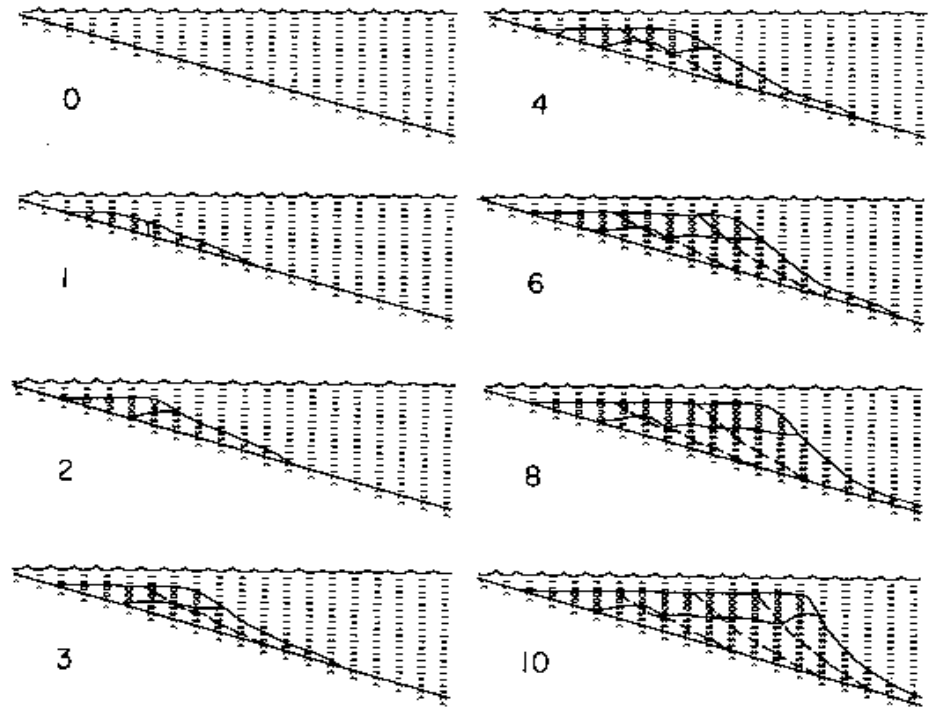

*Figure 2: Output of historical stratigraphic forward modeling program by Harbaugh and Bonham-Carter.*

The model produces sigmoidally shaped deposits that strongly resemble those formed on continental shelves. The model also can treat several sediment types at a time, by assigning each type a specific value of k. The proportion of each type is specified for input during each time. A mixture of sediment types is thus available for deposition in each cell.

Various authors deal with models of meandering streams. Particularly notable is the work of Bridge (1975). His model that utilizes widely accepted empirical formulas that relate a river's hydraulic regime with the geometry of its meanders. The model contains modules that simulate the plan form of meanders, the shapes of channel cross sections, the nature and occurrence of cut-offs (which are treated probabilistically), discharge during seasonal high-water periods, and long-term aggradation. Valley slope, discharge, and sediment characteristics are provided as input data to the computer program. Output consists of vertical cross sections, that can be directly compared with sedimentary sequences. The model is particularly useful in interpreting the sedimentary sequences produced by combinations of channel migration, aggradation, and seasonal discharge variations.

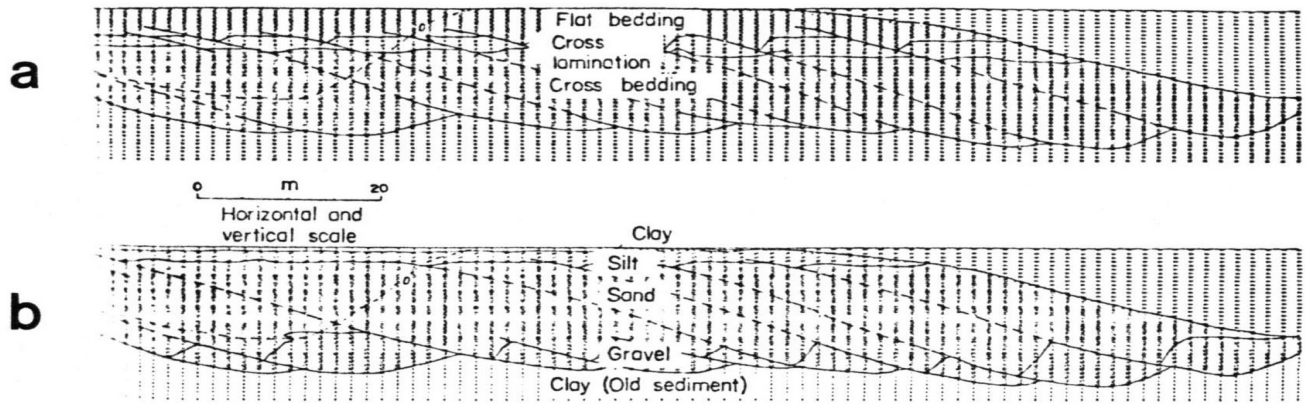

*Figure 3: Output of meander modeling program by Bridge.*

From the 1980's to today, increasing computer power and powerful algorithms propelled stratigraphic forward modeling methods into the realm of commercial applications for use in hydrocarbon exploration and production (Tetzlaff and Harbaugh, 1989). These include Geologic Process Modeler (GPM, from Schlumberger), Dionisos Flow (from Beicip Franlab), SedSim (from Stratamod), as well as non-commercial packages developed at universities such as Delft3D (from the University of Delft), and Carb3D (from the University of Bristol).

In the 2010's and 2020's the difficulty that some of these models have for honoring data promoted research to combine them with pattern recognition, artificial intelligence and machine learning methods. Some of these methods require many realizations to properly "train" a model to yield results that adequately reflect the true variability of real stratigraphic sequences. Furthermore, each of the many training results is often required to be complete (namely contain no missing data) a feat that is difficult with real-case scenarios and field measurements. Thus arose the need for a program that runs faster than most of the major packages, that contains tractable and open algorithms (rather than proprietary "black boxes"), even at the expense of some complexity, and that is accessible to researchers and students. That is the objective of SedSimple.

# Basic assumptions

## Grids

SedSimple utilizes grids of cells that are square in plan view and are arranged in layers that have variable vertical dimension. The vertical dimension is expressed as a z-value (elevation above an arbitrary datum) at the top corner of each cell.

*Node-centered vs. cell-centered magnitudes*

The properties of each cell (such as sedimentary composition) are also assumed to be represented at each node (cell corner) and vary bilinearly within the area of the cell. The properties are thus said to be node-centered (as opposed to cell-centered) in plan. In cross section however, the properties are assumed to be "tie-centered", meaning that a vertical line within a cell represents a single value, which is a vertical average of the property along the line. These concepts, while relatively simple, must be considered when displaying the output of SedSimple with graphic programs. They are further explained below. The graphic program provided with SedSimple (called Sirius) follows the same conventions as the simulation program to ensure consistent display.

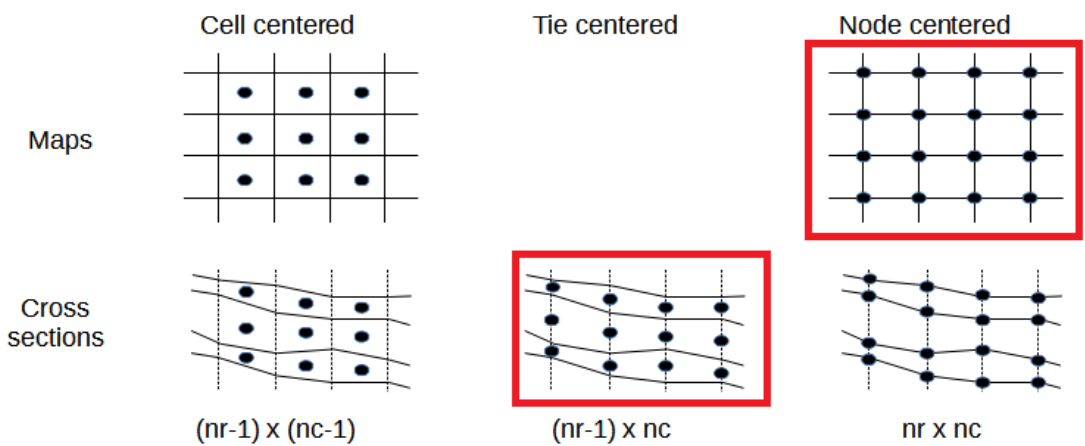

*Figure 4: Cell-, tie-, and node-centered data in maps and cross sections. Red frames show scheme used by SedSimple.*

When displaying a surface (such as a formation top) or a vertical section, property values (such as lithologies or petrophysical properties) are usually shown in color. The color may be constant across the cell, or it may vary gradually. The value that the color represents may represent an average for the cell, or the value at a corner (or "node") or, in the case of vertical sections, the value at the vertical boundary between cells.

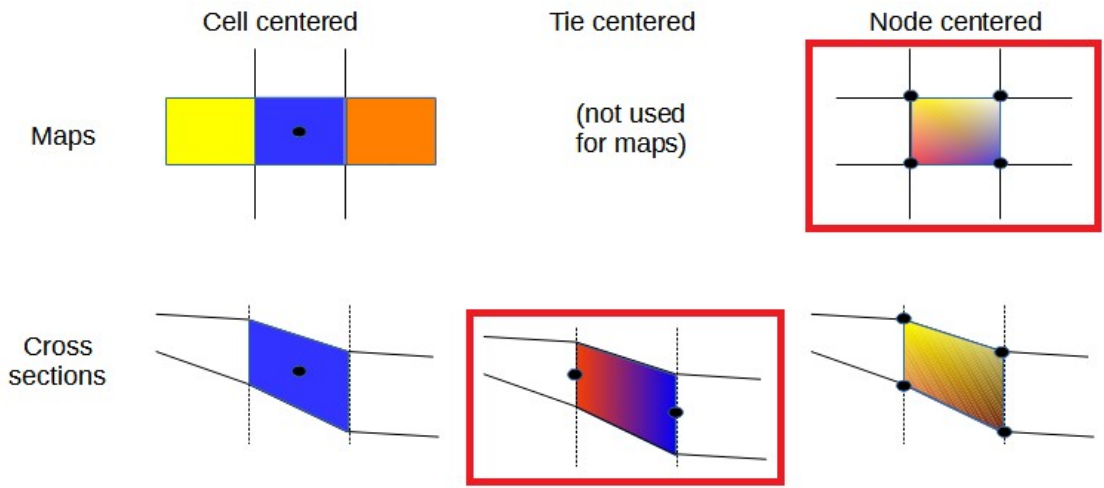

*Figure 5: Cell-, tie-, and node-centered data as they appear in graphics programs. Red frames show display preferred for use by SedSimple.*

It is important that the coloring scheme of cells both in plan as well as in section match the cell data scheme of the numerical algorithm. If, for example, the algorithm stores elevation data and properties at cell nodes (as SedSimple does), but the graphic program stores elevations at nodes and properties at cell centers (as many geologic display programs do), then either the information will be shifted by half a cell or it will be necessary to perform an average of cell corners before display. The companion program to SedSimple (called Sirius) ensures that the display data scheme is the same as that of the algorithm.

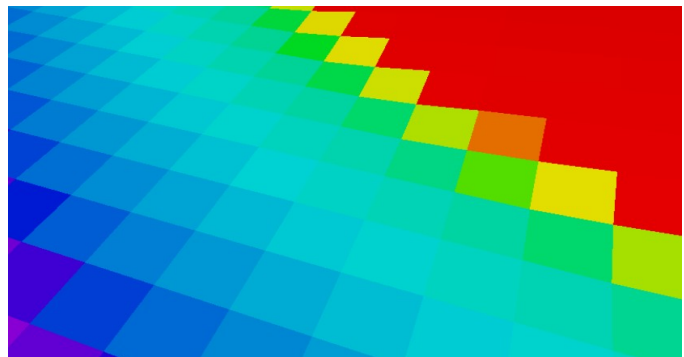

*Figure 6: Example of cell-centered display in plan, which is less than ideal for representing SedSimple data.*

However, if a node-center display package is used, care must be taken to ensure that color interpolation within a cell is adequate. Due to the intricacies of graphic display algorithms and their requirements to use triangles as primitives for graphic display, it is common that square or rectangular cells favor one diagonal over the other for intra cell interpolation. When values are highly variable but statistically isotropic, this may result in spurious features trending, for example, southwest-northeast, possibly leading to erroneous interpretations. This effect is shown in the next figure.

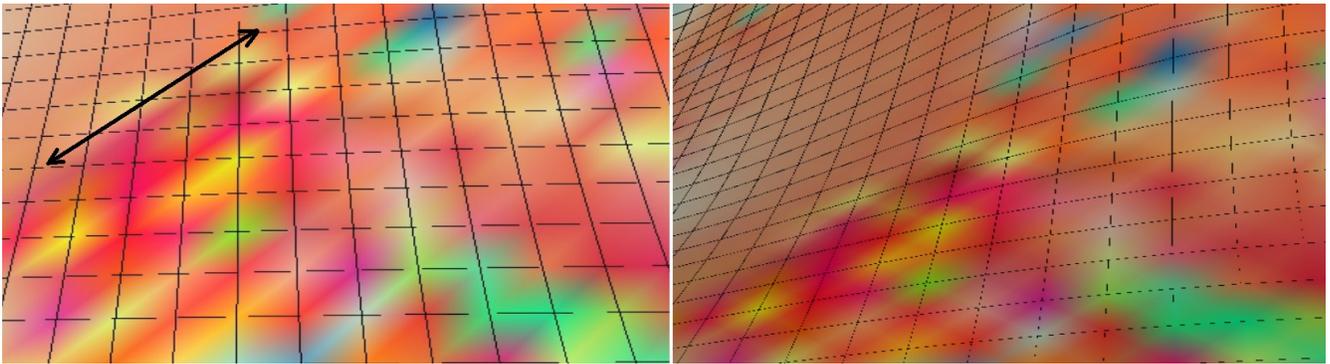

*Figure 7: Node centered display. Left: by program that shows diagonal smearing. Right: By program that does not show diagonal smearing.*

In the case of vertical sections, the cell-centered scheme for graphics often leads to apparent horizontal discontinuities, as shown in the next figure:

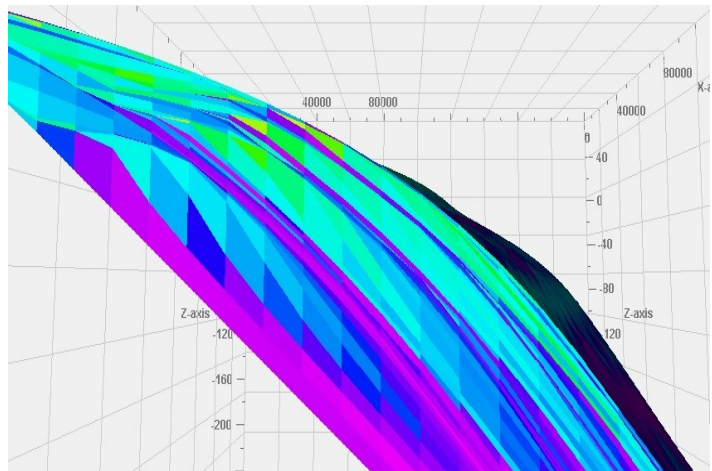

*Figure 8: Cross-sectional view of a program that shows cell-centered data, not used by SedSimple.*

The tie-centered scheme may be better suited for display of stratigraphic sequences in that it shows vertical discontinuities (which are often due to environment changes, without necessarily implying unconformities):

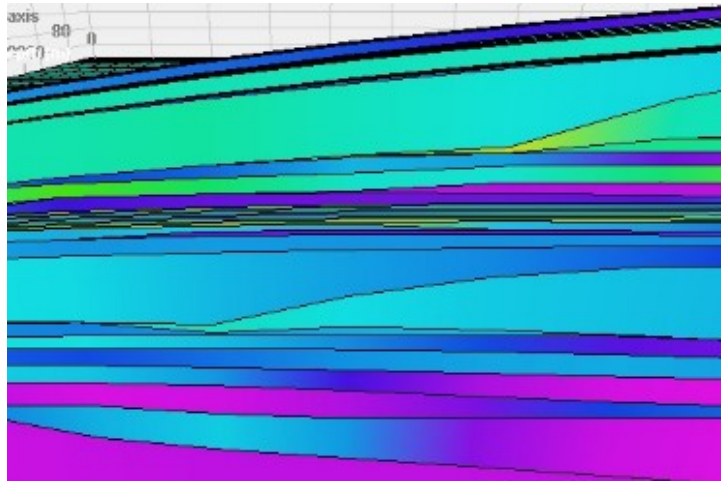
*Figure 9: Cross-sectional view of a program that displays tie-centered data. This scheme is used by SedSimple.*

Finally, there are also graphic display programs that show vertical sections using a node centered scheme:

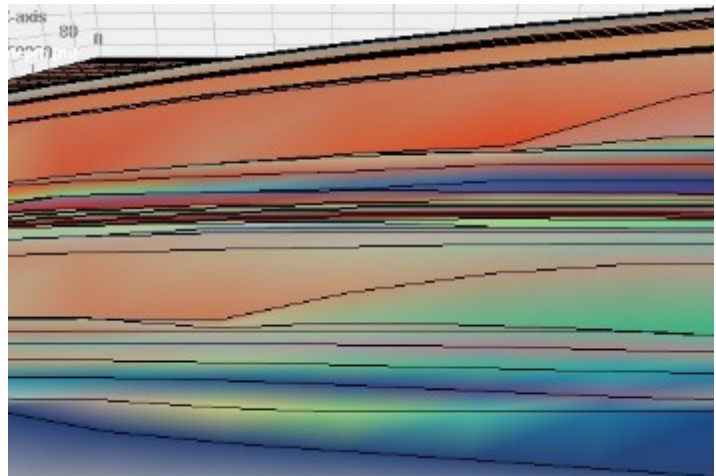
*Figure 10: Cross-sectional view of a program that displays node-centered data, not used by SedSimple.*

This may lead to excessive vertical continuity. If the information is actually tie centered, then the values represented by a color in one layer may "bleed" into the next adjacent (higher or lower) layer, also possibly causing interpretation problems.

*Erosion representation*

When representing layers in vertical cross sections, it is possible that a layer bottom truncates a lower layer due to erosion. Many graphic programs are not able to draw lines that are truncated within a cell. They force every top node to be lower than a younger layer. This may seem necessary and trivial but it can lead to errors in interpretation when visualizing the section, as shown in the next two figures.

The first figure shows a sedimentary sequence deposited until a hypothetical time *t*.

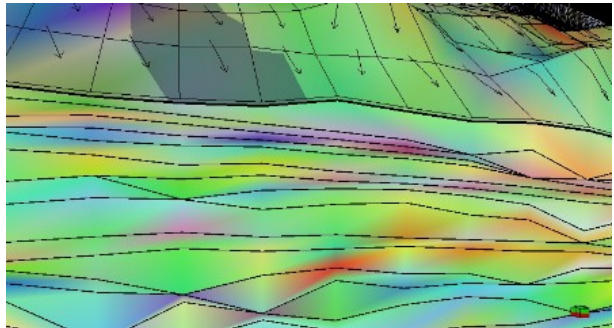

*Figure 11: Cross section of a sequence at time t, about to be eroded by a channel.*

Then at time t+1 erosion occurs, a channel is eroded, and another layer (of the same age as the channel fill) is deposited. If the algorithm now forces all lower nodes to not exceed the eroded surface (left side of the next figure), the shape of all underlying eroded surfaces will change, giving the impression that the channel started much earlier. If, instead, the erosion leaves underlying surfaces intact (namely eliminating the outline of the eroded portion but leaving the shape of the original layers intact), then the display is more likely to be interpreted correctly.

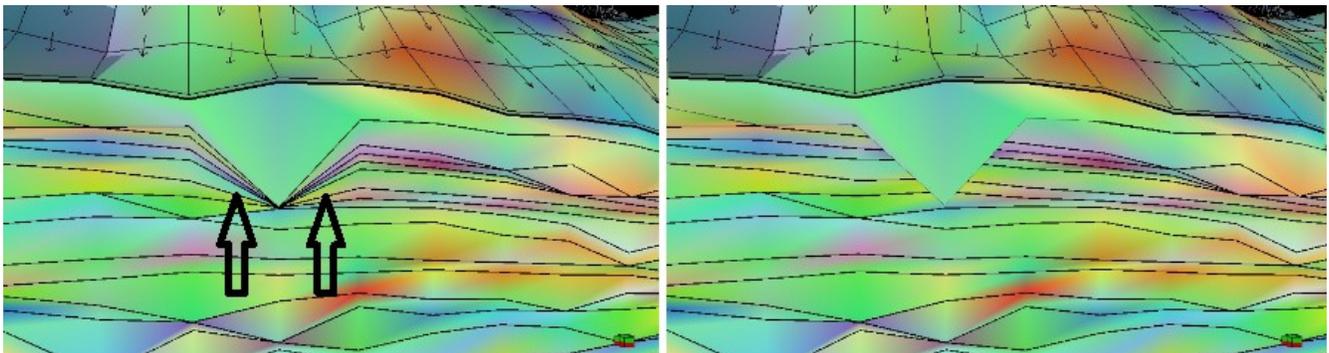

*Figure 12: Left: Cross section after channel erosion and deposition by a program that does not handle truncation. Right: Same display by a program that does handle truncation.*

## Time handling

SedSimple starts a simulation run by reading an initial state of the system from one or more files. For example, part of the initial state may be the initial configuration (topography or bathymetry) of the basin. It also reads the external input that is predetermined, such as sea-level change and tectonics. The details of these files are not explained here, but are contained in the user documentation.

While the program runs, it writes the current status of the model at several points in time called display intervals. When done the program writes the final state of the system.

In SedSimple any of these files (the intermediate outputs or the final output) can be readily reused as input to a new run (with the possible addition of suitable new parameters to specify new inputs and boundary conditions). This is important because if the program produces undesirable results beyond a certain time, it can be continued with different parameters without the need to perform the entire run from the beginning. Additionally, this feature makes it easy to develop models in which the processes change substantially throughout the simulation, keeping each set of parameters in separate files, as opposed to preparing complex files in which processes or events need to be turned off or started in the middle of a run.

*Time-step organization*

Just like space is divided into grid cells, time in dynamic programs is divided into steps. In SedSimple, the largest time steps are regular intervals that divide the total time represented by the model. These are called display steps. At the end of each display step, the program:

1. Updates the output file to reflect the latest state of the model.

2. Produces a "time-surface" from the current top of the sedimentary sequence. This time surface becomes the top of the youngest layer. Its properties are averaged vertically and placed at the grid nodes corresponding to the top layer.

Some algorithms contained in SedSImple proceed in time steps that are smaller than display time steps. We call these "internal method steps". These steps are necessary to avoid instability and to improve accuracy. However, in SedSimple the user need not be concerned with any of these steps as they are adjusted automatically to ensure accuracy and stability. Only display steps must be specified by the user to control vertical resolution vs. computer runtime.

## Properties

As mentioned earlier, SedSimple stores properties at grid nodes. The most important properties are the proportions of various sediment types, but the program also stores other variables (such as wave energy) and is prepared to add others in future versions, or user-extended versions.

*Sediment types*

SedSimple works with a number of "sediment types", defined as follows:
- A sediment type is an aggregate of particles all of the same size, shape, and density
- Sediment types have many properties, but their transport characteristics can often be summarized by a single parameter: particle fall velocity in water
- The diffusion coefficient is often tied to the particle fall velocity on water.

SedSimple has no limits to the number of sediment types used. However, the companion graphic program (Sirius) is currently limited to four sediment types. This may seem small, the program keeps track of sediment mixtures in continuous proportions. Thus, just two sediment types (sand and shale, for example) can give rise to many different lithofacies (sand, shaly sand, sandy shale, shale). The addition of more types (such as two types of carbonates, for example) can yield many more lithofacies.

For each run, the user specifies the sediment types by their sediment particle diameter. The algorithm assumes that the particles are spheres of density 2.67 g/cm3. The algorithm translates this to fall velocity in water and then to transportability (the ease with which sediment is eroded and transported). If the user knows that the sediment density is different, or the particles are not spherical, then the diameter must be adjusted to an "equivalent diameter" that produces a similar fall velocity in water.

*Sediment transport*

Modeling multiple sediments is obviously far more complex than modeling a single sediment. When eroding and depositing multiple sediment types, a 3D record of sediment types must be kept. With multiple types it is necessary to calculate the erosion depth that will produce an amount of mixed material that can be carried by the flow. This process involves "consulting" the underlying layers, usually represented by a 3D array of values in a computer program, which is computationally expensive.

Sediment is usually assumed to be carried by a flow in two different ways:

1. As bed load, consisting of particles that are very near the bottom of the flow and repeatedly in contact with it. Particles in bed load can move by: rolling (turning with almost constant bottom contact), or by saltation (jumping off the bottom in repeated short leaps.

2. As suspended load, consisting of particles that are distributed throughout the depth of the flow (though at higher concentration near the bottom) and are suspended due to turbulence.

As long as the total amount of each sediment eroded, transported, and deposited is correct, the mode of transportation is not important (at least for low-resolution stratigraphic purposes). Sediment load is often measured as grain volume per total fluid volume (including grains). This is the *volumetric* sediment load. When multiplied by flow velocity, it yields 3D volumetric sediment flux, which is the volume of sediment passing through a unit area perpendicular to the flow per unit time. The units of 3D volumetric sediment flux are therefore distance over time.

The most important measure of sediment transport is the sediment flux (whether 2D or 3D). A flow that is fast but carries a small load may have the same sediment flux as a flow that is slow with a high load (as the flux is the product of load times velocity).

There are many criteria that have been developed to estimate sediment flux. They are usually semiempirical and vary greatly form one another depending on whether they apply to a single sediment, a sediment mixture, or a single sediment within a mixture. They also vary depending on the conditions of the experiment. (For a sampling of these criteria, refer to Tetzlaff, 1989).

*"Equivalent load" principle*

A very useful principle that summarizes and simplifies the various sediment transport criteria is the "Equivalent load" principle. It establishes a way to "convert" a volume of any sediment mixture to an equivalent volume of a single hypothetical sediment type that will be carried under the same conditions. For example if we have sand and silt, a certain flow may be able to carry only 0.1 times the amount of sand than the amount of silt. If we divide the load of sand by 0.1 and the load of silt by 1, then the flow will be able to carry equal amounts of each "equivalent" sediment. This can be represented by the following expression:

$$\sum_{i=1}^{n} \frac{f_i}{c_i} \quad \text{Eq. (1)}$$

where:  n is the number of sediment types,
 $f_i$ are the seadiment fractions
 $c_i$ are the transportability coefficients.

The factor by which to divide the load is the sediment transportability. It is similar to the diffusion coefficient, and usually assumed to be dependent on the particle's fall velocity in water. The "equivalent load" of a mixture is the load of each type divided by its transportability. SedSimple makes use of this concept for all its modeled processes.

## Parallelization

Since forward modeling programs consume a lot of computer run time, it is natural to attempt to parallelize them. SedSimple currently does not contain any code that takes advantage of parallel processing. However, many of the algorithms contain steps that are independent of the result of previous steps, and are thus parallelizable.

When performing more than one run (as for example when trying the effect of gradually changing the input), most operating systems will take advantage of all present processors, and run different instances in parallel. However, lower-level parallelization requires suitable changes to the code. Thanks to the open-source concept, the knowledgeable user can taylor the program to multiple processors as desired.

# Processes modeled

## Sea-level change

Sea level is one of the most important factors controlling sedimentary processes. SedSimple uses many algorithms that depend on sea level. Sea level is entered as a time-dependent curve used by all processes that need it. The curve is referred to the level "zero" of the grid utilized as the initial topography.

The user specifies the sea-level curve in the form of a two-column entry to indicate time and sea-level elevation respectively. The times need not be equally spaced, but must be ascending.

## Diffusion

Perhaps the simplest way to model sediment movement is to assume that it moves down slope in proportion to the steepness of the slope, the characteristics of the sediment, and the environmental conditions such as water turbulence. This is not always strictly true. Sediment can move uphill carried by winds, ocean currents, other water currents, or biological activity. But in first approximation, it is a reasonable assumption. Furthermore, it is also common sense that the rate at which sediment moves is roughly proportional to the slope and is also affected by many other factors, such as the medium in which it is immersed, the process that causes the movement and the type of sediment.

The terms "dispersion" and "diffusion" are often used interchangeably in geologic modeling. In some disciplines (such as physics and chemistry) "diffusion" seems to have a more restricted meaning: it pertains to the movement of a substance in the direction opposite to the gradient of the same substance ("down slope"). Consequently, it also refers to the reduction of concentration at concentration "highs" (negative second derivative) and, conversely, increment at concentration "lows". In geology, this applies well to topography when we have a single sediment type, as sediment moves from high to low topographic elevation. However, when multiple sediment types are present, the term "diffusion" may be confusing as it might pertain to the "dilution" of high concentrations of one type of sediment regardless of topography. Nevertheless, the term "diffusion" is wide spread in the literature when referring to superficial movement of multiple sediment types, so we continue to use it throughout this paper.

### *Single-sediment diffusion*

SedSimple models diffusion of multiple sediment types simultaneously. However, to introduce the concept of diffusion we start by defining it for a single sediment. Modeling the diffusion with a model containing a single type of sediment is perhaps the simplest possible three-dimensional way of stratigraphic forward modeling. While it may not be very applicable to real cases it is very instructive as a first step to develop more complex methods.

The following is the basic equation for single-sediment diffusion. It simply says that topographic highs will be eroded and lows will be filled:

$$\frac{\partial z}{\partial t} = k \nabla^2 z + s_n \qquad \text{Eq.(2)}$$

where:
  $z$ = topographic elevation
  $k$ = diffusion coefficient
  $t$ = time
  $\nabla^2 z$ = Laplacian of $z$
  $s_n$ = sediment source term

As a consequence of diffusion, highs are eroded and lows are filled, as schematically shown in the next figure:

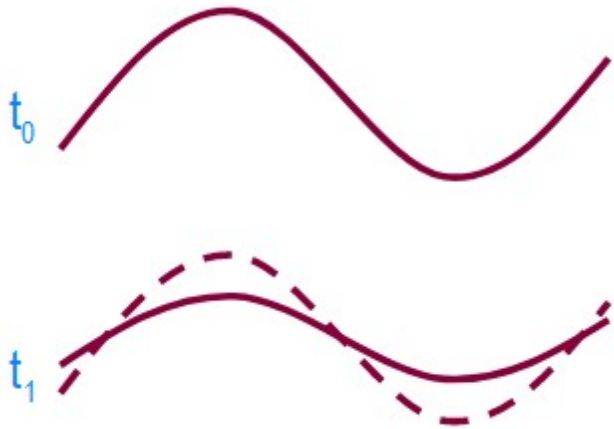

*Figure 13: Single-sediment diffusion and schematic evolution of topography.*

The diffusion coefficient *k* determines how "fast" this process will happen. It will depend, among other variables from the sediment type and the "energy" of the environment, and can be estimated from field measurements (Simon et al., 2022). The dependency on the "energy" of the environment (actually a measure of turbulence), can be very strong. Over long periods of geologic time, coastal areas have the highest "energy". Therefore, it is common to assume a depth-dependent diffusion coefficient curve (Kaufman, 1991), with the highest coefficient near sea-level:

- Dispersion coefficient varies with depth (elevation) relative to sea level

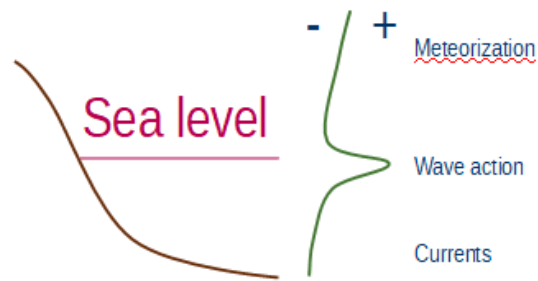

- Dispersion coefficient is different for each sediment type (gravel, sand, silt, clay)

*Figure 14: Schematic curve that determines diffusion coefficient as a function of depth.*

This assumption has a strong effect on the resulting processes. Under the assumption of a spatially constant coefficient, diffusion leads to relatively featureless surfaces, whereas with a coefficient that varies vertically and is higher at sea level, more realistic shoreline profiles are produced. Schematically this is as follows:

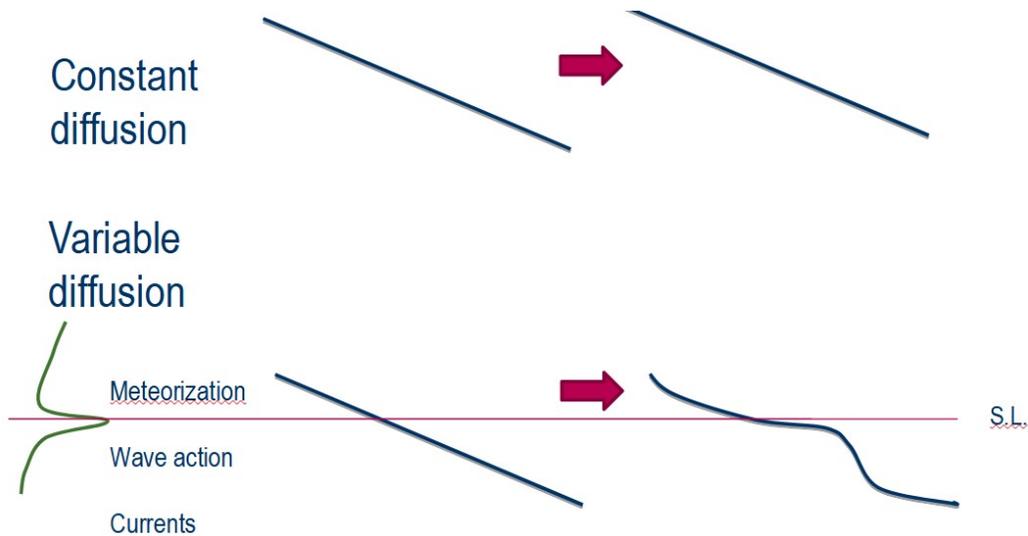

*Figure 15: Schematic cross section through a diffusion model in which sediment is entered at the high end and the same amount is withdrawn at the lower end, after reaching steady state. Top: Constant diffusion comparison. Bottom: diffusion coefficient varying with depth.*

The diffusion coefficient has units of surface over time. This is explained by the next figure:

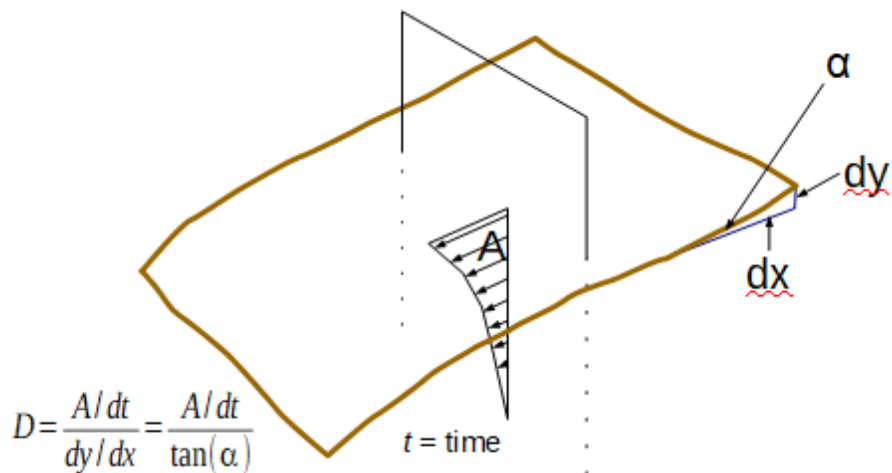

*Figure 16: Interpretation of diffusion coefficient.*

In any area perpendicular to the flow, there will be a volume of sediment passing through it per unit time. This is the 3D volumetric sediment flow, and has units of volume over time (not to be confused with the 3D volumetric sediment flux, defined later in this chapter). If the flow is divided by the width of the area, we obtain the 2D volumetric sediment flux, which has units of surface over time. Another way to define the 2D volumetric sediment flux is the area of sediment that goes through a vertical line per unit time.

The 2D flux is assumed to be proportional to the slope (tangent of the slope angle) to the "energy" of the environment, and to characteristics of the sediment (often a function of the particle's terminal fall velocity in water).

*Stability*

An important issue with most numerical methods is stability. If the time step is made too large, the algorithm will produce incorrect (erratic, oscillating, and eventually infinite) results. Fortunately, there are criteria that, if properly applied, prevent this from happening. The main criterion for this simple finite-difference method outlined above is the following:

$$\frac{D \, \Delta T}{(\Delta X)^2} \leq 0.25 \qquad \text{Eq. (3)}$$

where:
$D$ = Diffusion coefficient
$\Delta T$ = Method time increment
$\Delta X$ = cell side

The application of this criterion is simple, but gets more complex when multiple sediments are considered (as will be shown in the next chapter). SedSimple incorporates this principle in a way that is completely transparent to the user. As mentioned earlier, the user only needs to specify the display time interval and the program will figure out the internal algorithm step sizes to ensure stability.

*Verification of the diffusion principle in nature*

A consequence of diffusion using a depth-dependent coefficient is that the depth-derivative of the equilibrium profile (i.e. the stable profile when the same amount of sediment is input at the high end as is withdrawn at the low end) should be similar to the vertical diffusion function. This is not demonstrated here, but is easily shown through the use of differential equations.

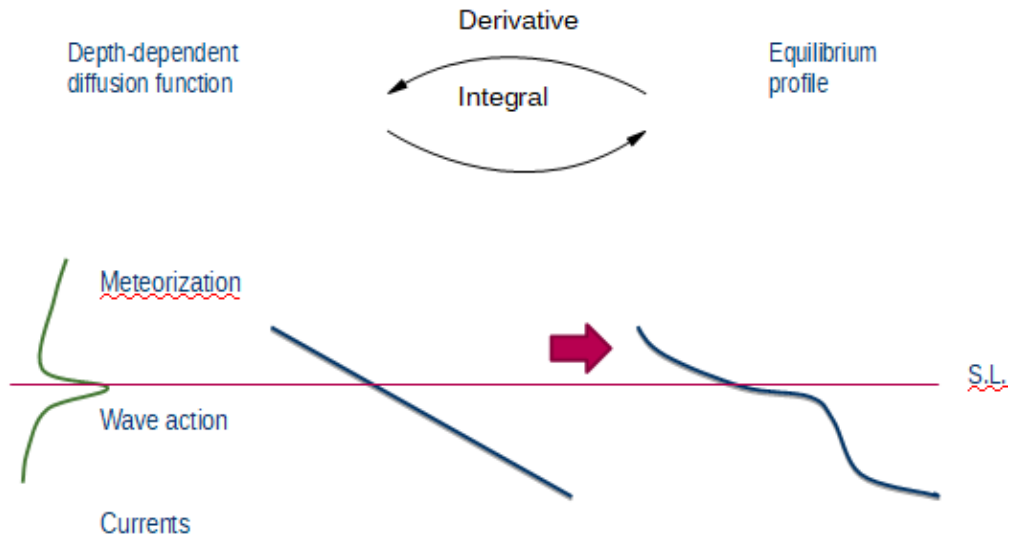

*Figure 17: In a steady state profile (same amount of sediment introduced at left is removed at right), the vertical profile is the vertical derivative of the diffusion coefficient.*

There is a remarkable similarity between the schematic profile in the previous figure and a graph of world-wide, or continental-scale depth and elevation data:

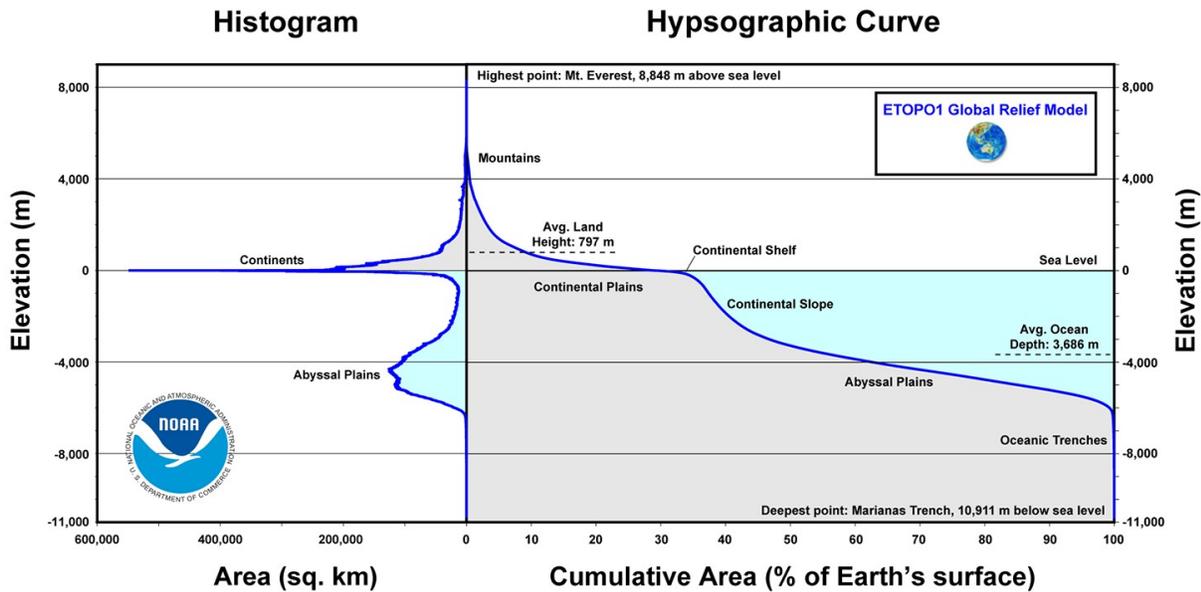

*Figure 18: Statistical distribution of global elevations (courtesy of NOAA)*

The upper part of this graph (down to the continental slope) may be considered approximately in equilibrium as sediment is input from erosional processes, and ultimately ends a=in the abyssal plain. The "energy" (diffusion coefficient) is maximum on the shelf due to wave action. The upper part of the histogram (left side of the figure above) also mimics the diffusion coefficient.

The lower part of the graph (abyssal plain and trenches) does not follow this scheme because it is not in equilibrium (sediment constantly accumulates at depth) and also because hypsographic curve may be controlled further by crustal processes (oceanic vs. continental crust).

While this graph is not "proof" that the diffusion principle works, it shows that it is at least roughly compatible with worldwide observations and may be used for large-scale models.

For areas vertically within 100 m of current sea level (next figure) we see a small "indentation" in the profile which represents the higher coastal energy after the last sea-level rise (probably less than 10,000 years).

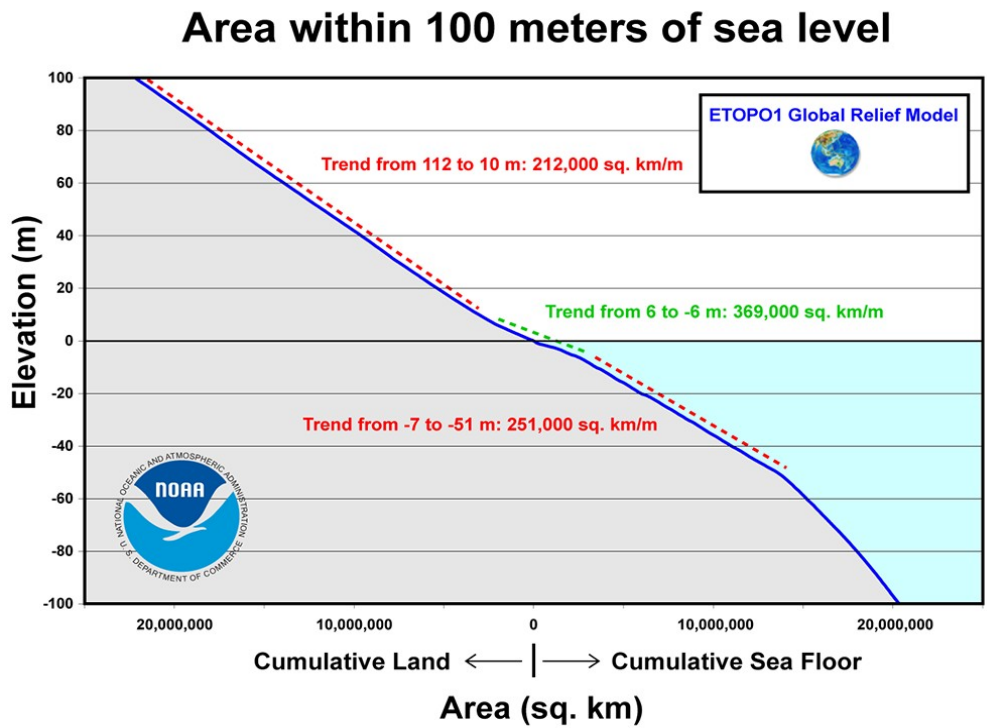

*Figure 19: Detail of statistical distribution of global elevations (courtesy of NOAA).*

*Multiple sediment diffusion*

The mathematical expressions for multiple sediment diffusion are necessarily different for erosion and deposition, because deposition is not just the reciprocal of erosion. Erosion must remove sediments in the same proportions as they are present in the op layer, whereas deposition may favor coarse sediments in the load, depositing them first and resulting in sorting.

Erosion occurs when the flow conditions are such that the "equivalent" sediment at the bottom will be eroded. Conversely, deposition occurs when the "equivalent" sediment being carried by the flow is deposited:

Erosion:

$$\frac{dz}{dt} = \nabla^2 z \frac{1}{\sum_{i=1}^{n} \frac{F_i}{c_i}} \quad \text{Eq. (4)}$$

When erosion occurs, all sediment is eroded down to depth *z*.

Deposition:

$$\frac{dz}{dt} = \nabla^2 z \sum_{i=1}^{n} \frac{dz_i}{dt} = \nabla^2 z \sum_{i=1}^{n} \frac{f_i}{c_i} \quad \text{Eq. (5)}$$

When deposition occurs, coarse sediments are assumed to be deposited first.

*Numerical method for multiple-sediment diffusion*

Here we describe briefly the numerical method used by SedSimple to model multiple-sediment diffusion. It is not the same method used by major commercial programs. It has the great advantage that it easily ensures sediment continuity (conservation of mass for each sediment type) and its stability can be studied and controlled with relative ease and it does not require the use of an "active layer" (a hypothetical layer in which all movement takes place). It proceeds roughly as follows:

1. Work on the nodes (cell corners) of a square grid.

2. Select those nodes that are surrounded only by "down-slope" (or horizontal) ties. These are the local maxima (nodes surrounded by lower or equal-height nodes). "Process" these nodes (what "process" means is explained later).

3. Proceed down slope: "Process" nodes whose upslope neighbors are already processed, until all nodes are processed.

To "process" a node means:

1. Determine whether there is erosion or deposition at that node, based on the second derivative.

2. If there is erosion, erode the amount of equivalent sediment, and move this amount down slope, prorated to each down-slope tie according to its slope (tangent of slope angle). The amount moved is kept on hold at each lower node.

3. If there is deposition, deposit the equivalent amount, coarse first, using the sediment "on hold" at that node. If there is no more sediment "on hold" stop depositing. If there is sediment left over, move this amount down slope, prorated to each down-slope tie according to its slope.

*Example of diffusion model*

Below we present one example of SedSimple output using diffusion and sea level change. The colors represent fout sediment types (red: coarse sand, green: fine sand, blue: silt, and black: clay). Additive combinations of these colors are used to represent sediment mixtures in all proportions (for example yellow is an additive combination of red and green, representing coarse sand and fine sand).
The run involves 2 million years and 100 layers. Run times shown in the figure captions correspond to total program runtime (including reading input files and writing intermediate and final results) on a modern laptop computer (Intel Xeon CPU at 3.20 GHz with 64 GB of RAM).

The model starts with a uniform plane sloping 5% containing equal proportions of all sediment types.

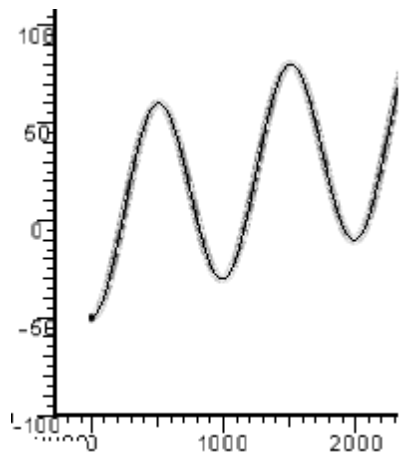

*Figure 20: Sea-level curve for next example: x-axis: thousands of years, y axis: meters.*

The system is closed (all sediment comes from the erosion of underlying sediment and is not allowed to exit at the border. The result is shown in the following figure:

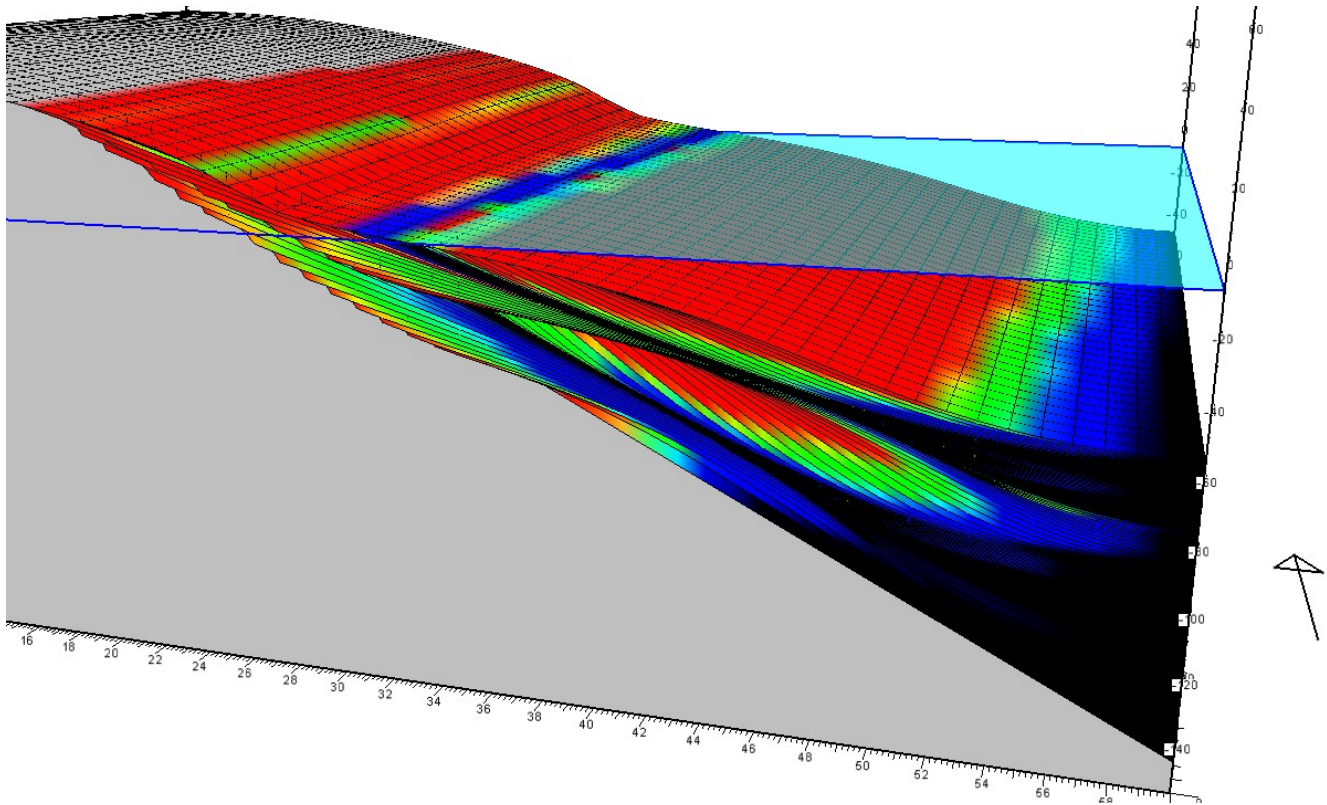

*Figure 21: Example of diffusion model with 2 sea-level cycles showing 100 layers. Runtime = 8.6 seconds.*

## Channels, fans and deltas

Traditional methods to model free-surface flow coupled with erosion, transport and deposition of sediments in natural channels for stratigraphic purposes include finite difference methods for steady flow and particle methods for unsteady flow. SedSimple uses neither of these methods. Instead it utilizes a method (that we have called "whole-channel modeling"), that combines the effects of flow and transport of multiple sediment types to model the origin, migration, and abandonment of channels under various paleoenvironmental conditions.

The whole-channel method follows the trajectory of a channel from the source to the point where the flow ends, either due to dissipation in a basin or by reaching a local low that cannot be overcome by the flow. The method uses some concepts from earlier meandering river models (Ferguson, 1984, Sun et al., 1996) but extends them to include modeling the channel's longitudinal profile and cross section. The trajectory is represented by a continue curve in plan. Initially the curve follows the path of maximum descent. However, the channel then migrates sideways due to outward erosion and inner deposition at bends. The algorithm is based on the following formula:

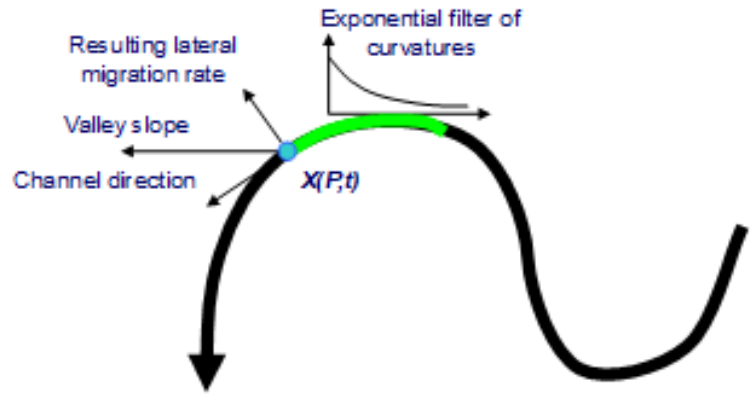

$$r = \int_0^P \left[ ck_c \exp(-q_c(P-p)) + (1-\cos(\alpha)) \cdot sign(\sin(\alpha))k_s \exp(-q_s(P-p)) \right] dp \qquad \text{Eq. (6)}$$

where:
$r$ = rate of lateral channel movement (positive to the right of channel direction)
$c$ = signed curvature (inverse of curvature radius, positive when turning left)
$kc$ = constant, amplitude of curvature effect
$qc$ = constant, rate of decay of curvature effect
$\alpha$ = angle between general valley slope and channel heading
$ks$ = constant, rate of decay of slope effect
$qs$ = constant, intensity of slope effect (dependent on valley slope angle)

As the channel flows, it shifts sideways at rate that depends on the immediately preceding curvature of the channel (and on several other parameters). The higher the curvature, the higher the rate of shifting. If the channel touches itself at a different point, avulsion occurs. The intermediate part of the channel is abandoned causing a sharp turn in the active channel that becomes a new meander. True meandering (in the sense of wide turns spanning more than 180 degrees) only occurs under certain conditions. If the slope is sufficiently high and the flow is relatively low, then the channel will only turn slightly, giving rise to a channel that turn only within a narrow range of directions and producing the pattern of a braided stream.

The width and cross section of the channel are determined by the flow velocity and depth (in turn calculated from semi empirical channel-flow formulas) and the proportions of the sediment types present at the bed and carried by the flow. The amount of sediment eroded, transported and deposited is calculated using the transport capacity concept previously described.

The user can specify a continuous flow rate and sediment supply, or may provide sporadic events of given frequency and duration.

*Fluvial, alluvial and deltaic*

Although the algorithm is the same for all kinds of channelized flow, the parameters are different for rivers, deltas and turbidites. The algorithm identifies a flow as either a river or a turbidite depending on the location of the source. The flow is assumed to be a river when the source is above sea level. When the river reaches sea level, the flow is assumed to become hypopycnal (namely, be less dense than the sea water). Therefore, it is no longer affected by the underlying bathymetry. Hypopycnal flow tends to slow down as if reaching a horizontal surface (the sea level) and gradually deposits all its load due to the steadily decreasing velocity (Burns and Meiburg, 2015).

While above sea level, if the topographic slope decreases suddenly, then it is likely that a very large portion of the sediment load will be deposited. This will typically give rise to an alluvial fan deposit. The deposition is not imposed arbitrarily, but arises naturally from local conditions, namely the flow becoming insufficient to carry the sediment load.

*Example of fluvial model*

The following figure shows an example of a fluvio-deltaic model with no sea-level change, run for a simulated time of 150 ka (kilo years) with 50 layers:

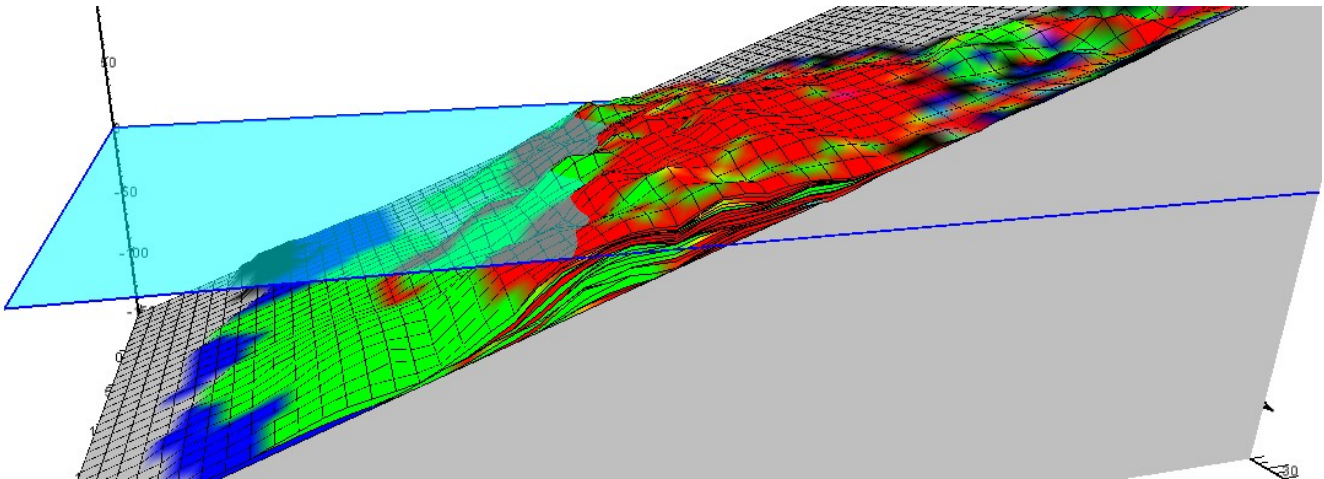

*Figure 22: Example of fluvio-deltaic model with no sea-level change. Runtime = 1.08 seconds.*

Another example of a fluvial model is shown under the section on "Aggradation and deep-sea sedimentation" later in this paper.

*Turbidites*

Numerous numerical models of turbidites of various degrees of complexity have been described in the literature (Meiburg et al., 2015, Nasr-Azadani et al., 2014, Parkinson, 2014, to name only a few of the models of interest for stratigraphic forward modeling). SedSimple uses essentially the same algorithm for whole-channel simulation. The algorithm assumes that the flow is turbiditic when the source is below sea level. The density of the flow has an important influence on the evolution of a turbidity system and its deposits (Fu et al., 2019). However, the exact density is hard to estimate from sediment load and flow velocity and depth. In SedSimple, flow density of turbidites is assumed to be fixed at a value higher than sea-water density. For lack of better local knowledge, the value is set in the code (typically at 1.5 g/cm$^3$). The flow continues along the sea bottom (possibly in a braided or meandering pattern) until slowing down enough to deposit all its sediment load, typically forming a lobe.

Turbidite deposits change downstream according to various factors (Heerema et al., 2020, Shimizu et al., 2017). In SedSimple downstream transitions from channelized flow, to distributary channels, lobes and distal deposits are caused by the mutual interaction of slope, flow velocity and sediment load, naturally evolving from source to basin floor.

*Example of turbidite model*

The following figure shows an example of a turbidite model run with SedSimple. It starts with a slope of 5% adjacent to a slope of 1% (respectively representing the continental slope and rise). The area has a total size of 6 km by 4 km, with cells 100 m a side. No sediment is input. All deposited sediment comes from the erosion of the preexisting surface. There are 120 flows each of which carries 10 million cubic meters of water.

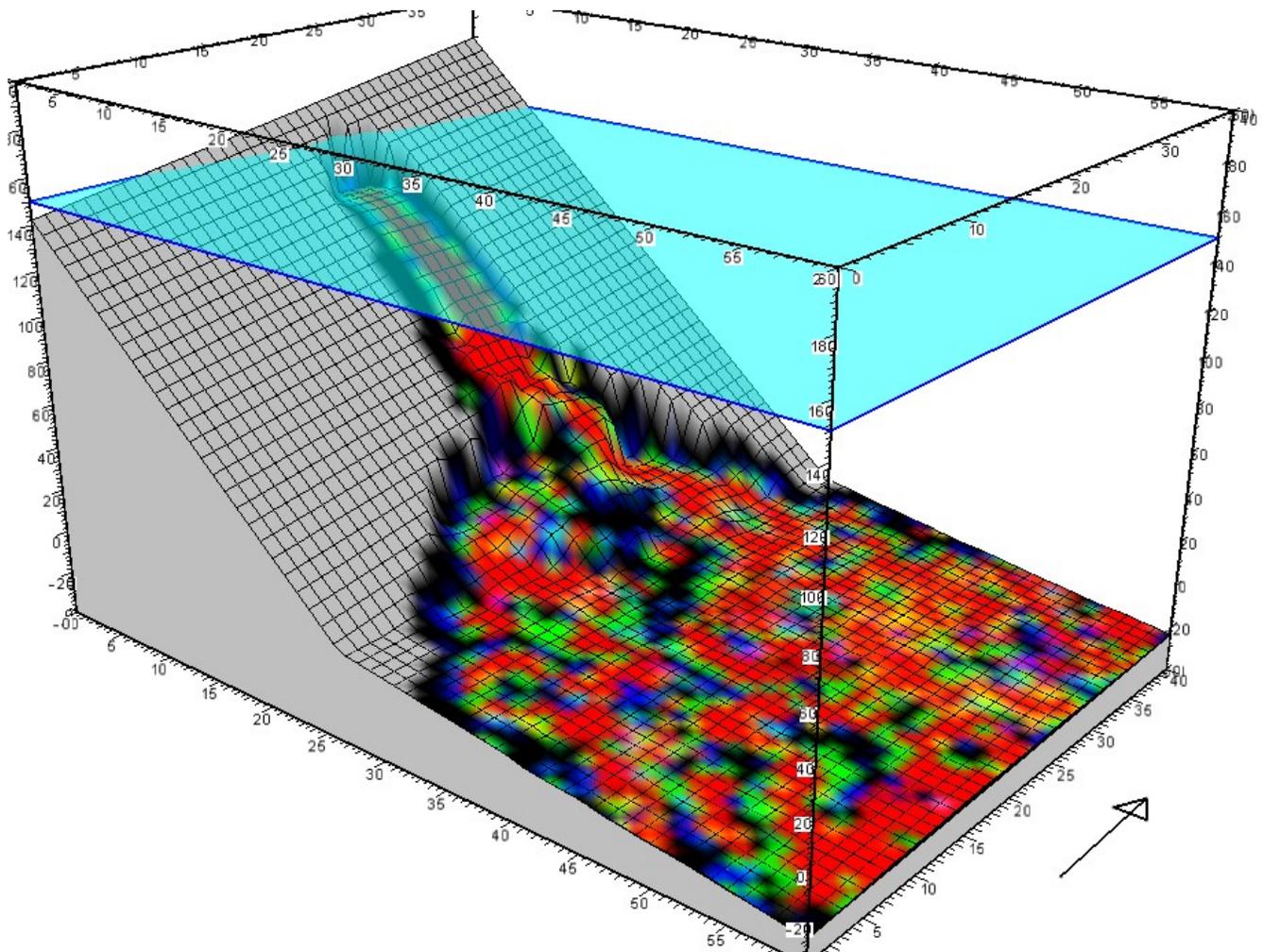

*Figure 23: Example of turbidite system representing 120 ka and 120 flows. Runtime = 0.63 seconds.*

A detailed transversal section is shown below:

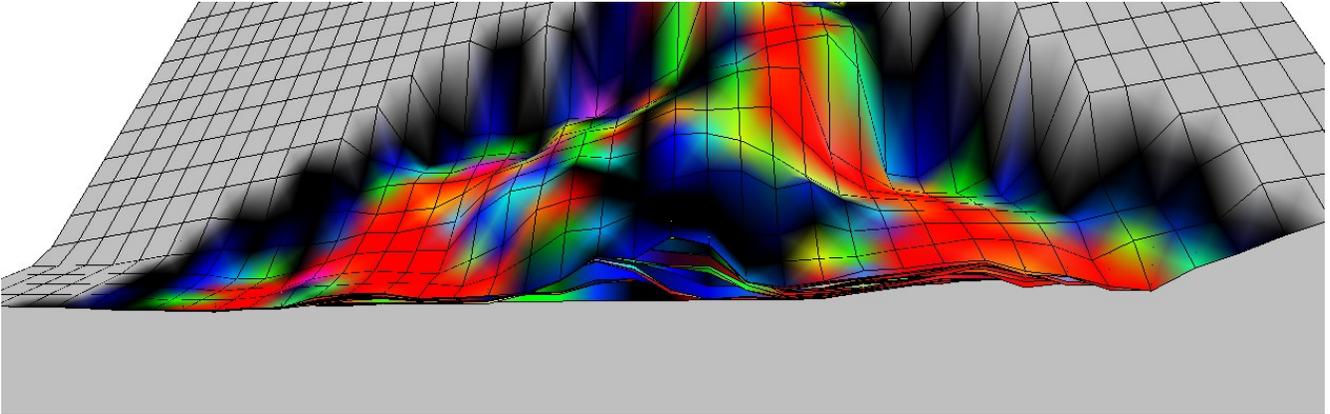

*Figure 24: Transversal section through turbidite model shown in previous figure.*

Further examples of a tubidite models is shown under the sections on "Aggradation and deep-sea sedimentation" and "Tectonics" later in this paper.

## Wave action

In general, wave modeling in stratigraphic forward models takes one of two possible forms:

1. Modeling the wave front and its progress through time, including its energy losses and ability to move sediment. This is the more rigorous approach and is adequate for modeling wave refraction and the effect of shallows and land masses. These methods in turn can be divided into:
    (a) Node-based algorithms., which rely on establishing the travel time of the wave front from the source to each node by visiting nodes (cell corners) that are adjacent to nodes that already have times assigned.
    (b) Front-based algorithms which model the wave fronts as continuous curves as they progress through time.
    This set of methods (1(a) and 1(b)) are usually very accurate but also consume significant computer resources (Jeschke and Wojtan, 2015).
2. Modeling the general wave direction and intensity (often derived from data or assumptions about the prevailing winds), and the general effect of waves on sediment as they approach shallow areas. This method is less rigorous but faster and easier to use.

SedSimple takes an intermediate approach between these two. It models the transport of weave energy along parallel lines that are perpendicular to the wave direction. These lines represent wave crests only in deep water, where the velocity is constant. When the waves reach shallow water, the wave energy is transferred laterally in order to model refraction, The process is explained in further detail below.

The first step in a model of wave propagation is to establish the wave celerity (a rough analog of "wave velocity"). It provides a means to estimate wave propagation and refraction.

Several well known formulas provide wave celerity. There are different formulas for "deep" and "shallow" water, as follows:

**Deep**
$$c = \frac{g}{2\pi} T = \sqrt{\frac{gL}{2\pi}} =$$
$$1.565 [m/s^2] T = 1.251 [\sqrt{m}/s] \sqrt{L} \quad (if \ d/L_D > 0.5) \quad \text{Eq. (7)}$$

**Transitional** $\quad (if \ 0.05 \leq d/L_D \leq 0.5) \quad$ Eq. (8)

**Shallow** $\quad c = \sqrt{gd} \quad (if \ d/L_D < 0.05) \quad$ Eq. (9)

where:
  $c$ = celerity
  $g$ = gravity
  $T$ = period
  $d$ = depth
  $L$ = wave length
  $L_D$ = deep-water wave length

These can be combined into a single "unified" formula, which is the only formula used by SedSimple:

$$c = \sqrt{\frac{g}{\kappa} \tanh(\kappa\, d)} \quad \text{Eq. (10)}$$

where:
  $c$ = celerity
  $g$ = gravity
  $d$ = depth
  $\kappa = 2\pi/L$
  $L$ = wave length

The latter formula approaches each of the preceding (deep and shallow) formulas, while also providing a suitable transition at intermediate depths.

The program proceeds in steps (labeled S1, S2, etc. in the next figure). These are not physical time steps, but simply computational steps. Each step corresponds to one of a set of parallel equidistant lines. The algorithm starts with the first line that intersects the area from the direction of wave provenance. This line is divided into a set of equal-length segments. These segments remain the same as long as the wave is in deep water. When the next line reaches shallow water the segment end points are moved laterally according to the lateral change in celerity to model refraction.

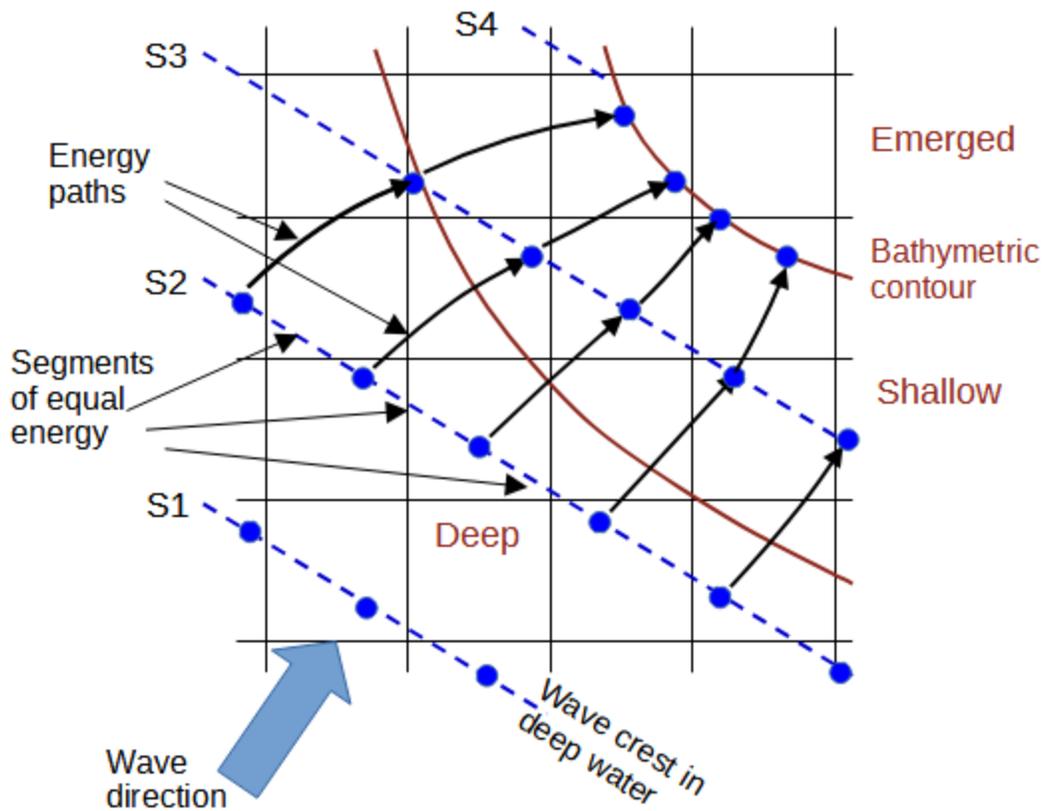

*Figure 25: Algorithm for transmission of wave energy.*

*Wave energy dissipation*

The effect of waves on the movement and mixing of underlying sediment is usually assumed to be controlled primarily by the energy dissipated by waves near the bottom (Liu et al., 2020). This quantity, often called "energy" or "energy of the environment is more precisely described in physical terms as "power dissipated per unit area".

There are plenty of measurements and information on the power that is transmitted by waves as they travel near the water surface, but there is much less information on the conditions and variables that affect power dissipation.

SedSimple assumes that energy dissipation takes two forms:
1. Energy dissipated by water viscosity during wave movement.
2. Energy dissipated by friction against the bottom

Sediment is assumed to move along the gradient of bottom-friction dissipation (i.e. from high bottom friction to low bottom friction).

## Example of wave-action model

The following figure shows an example of a wave-action model. It starts with a sloping surface with a mound (left side of the figure). After 30 ka of simulated time the mound is partially eroded and deposition occurs around it so that it coalesces with the shoreline. There is no sea-level change. A small amount of diffusion is also assumed in order to "feed" the shoreline from higher areas. This causes the large sand area (which is a very thin layer) just upward from the shore.

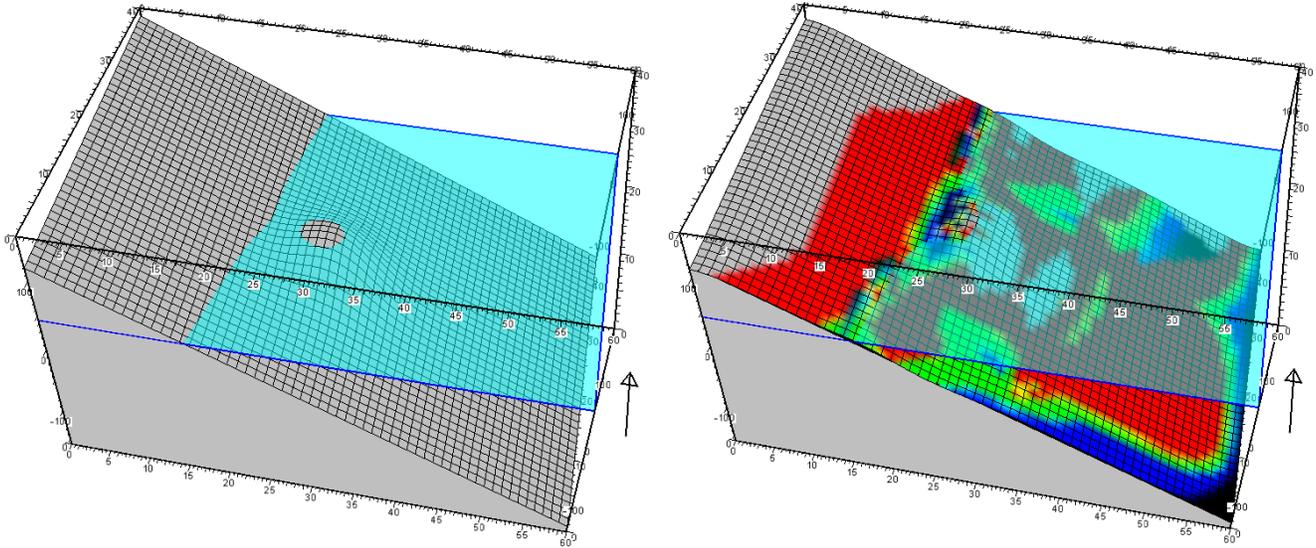

*Figure 26: Initial and final state of wave-action example. The mound is initially detached from the shoreline but coalesces with it due to erosion and deposition.. The model involves 10 layers and a simulated time of 30 ka. Runtime = 0.87 seconds.*

A detailed longitudinal cross section is shown below:

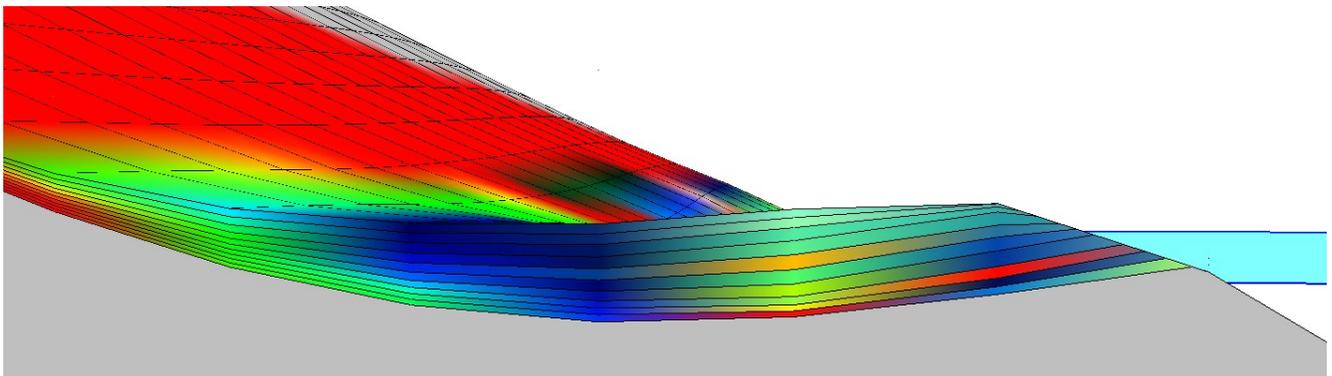

*Figure 27: Longitudinal cross section through center of wave model shown in previous figure.*

## Carbonates

Carbonates grow due to chemical and biological phenomena controlled by paleo environmental conditions. Once they are formed, they can be eroded, transported and deposited just like siliciclastic sediments. The combination of carbonate growth and redistribution in a single model can reproduce many processes (such as allocyclic and autocyclic processsses) that would not be possible to model with separate models (Duan et al., 2000, Hill et al., 2012).

The following figure shows, schematically, the factors for carbonate growth in SedSimple, namely Wave action (energy dissipates by bottom friction) and water depth (which encompasses the ability of light to reach the bottom). Other factors (such as water turbidity, temperature, and salinity) are not considered separately, but can often be encompassed into the other factors.

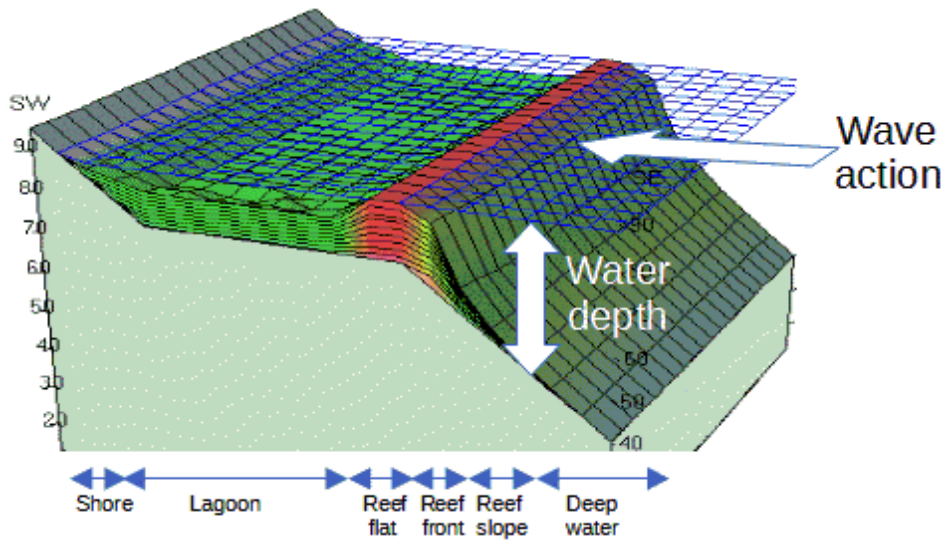

*Figure 28: Factors affecting carbonate growth.*

Modeling carbonate growth consists of:

1. Obtaining a measure of the growth factor (for example water depth, wave energy, location on production map)
2. Applying this factor to determine growth using a function that ties the magnitude of the factor to a growth rate

*Growth rate functions*

The effect of depth and wave action is specified by the use of production curves. For each carbonate sediment type, two curves may be provided: One that specifies production as a function of depth, and the other as a function of wave energy.

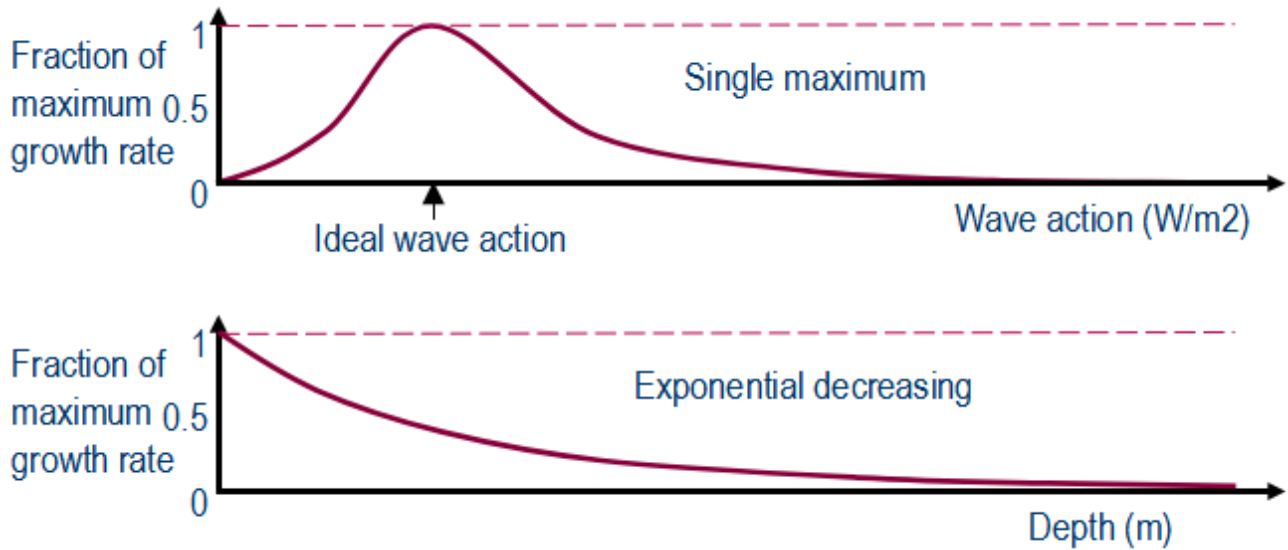
*Figure 29: Examples of carbonate growth curves.*

These curves normally vary between 0 and 1, and represent a fraction of the maximum production for that type. The maximum production in turn is specified for that type as a value of mm/a (millimeters per year). The curves are arbitrary and user-provided. Normally for wave action, a curve with a single maximum (representing the wave energy that provides ideal growth conditions) is used. For depth dependence, the most common curve will be exponential (decreasing), as many carbonates will grow ideally in shallow waters. However, it may have any shape, as certain carbonates (some corals for example) may grow best at a limited range of depths.

These functions define individual growth rates caused by each factor, but they do not specify how the growth factors should be combined when more than one factor is at work. SedSimple uses multiplication of both factor to determine growth rate, so if one factor is provided but is close to 0, the growth rate will be low.

In a few occasions in may be desirable to combine factors additively. For example, a certain carbonate may grow well either in shallow water or within a given range of wave energies. These cases are relatively rare as most carbonate growth requires a set of conditions, all of which must be met for favorable growth. For additive combination of factors, the user can still specify two separate growth criteria with the same carbonate type.

*Example of carbonate growth*

The following example shows a cabonate model that represents the formation of an atoll over a sea mound under rising sea level. There are two carbonate types:
1. Reef (blue color) that grows in response to moderate wave action.
2. Lagoonal and other (green color) that grows at shallow depths.

Waves enter the area from three directions: Predominant waves are from the east (to simulate a predominant wind direction), while minor waves enter the area form the southwest and northwest.

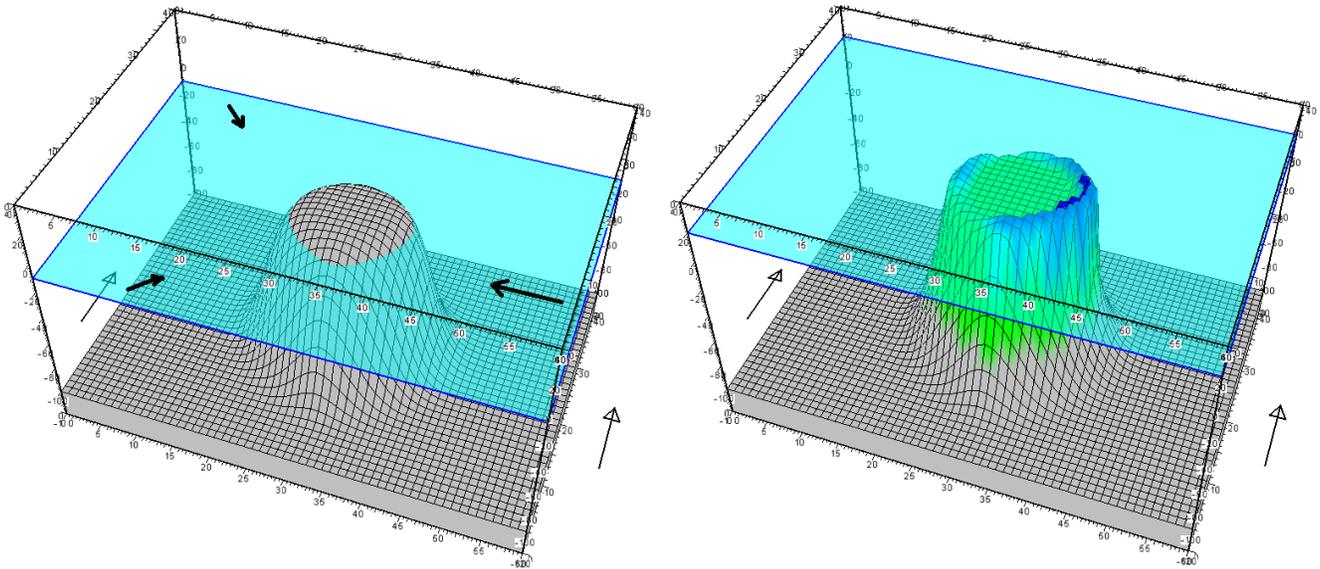

*Figure 30: Example of an atoll model through 30 ka simulated time and 30 layers with constantly rising sea level. Runtime = 3.67 seconds.*

As a result, reefs develop mainly on the eastern side. This is seen more clearly in the following detailed cross section:

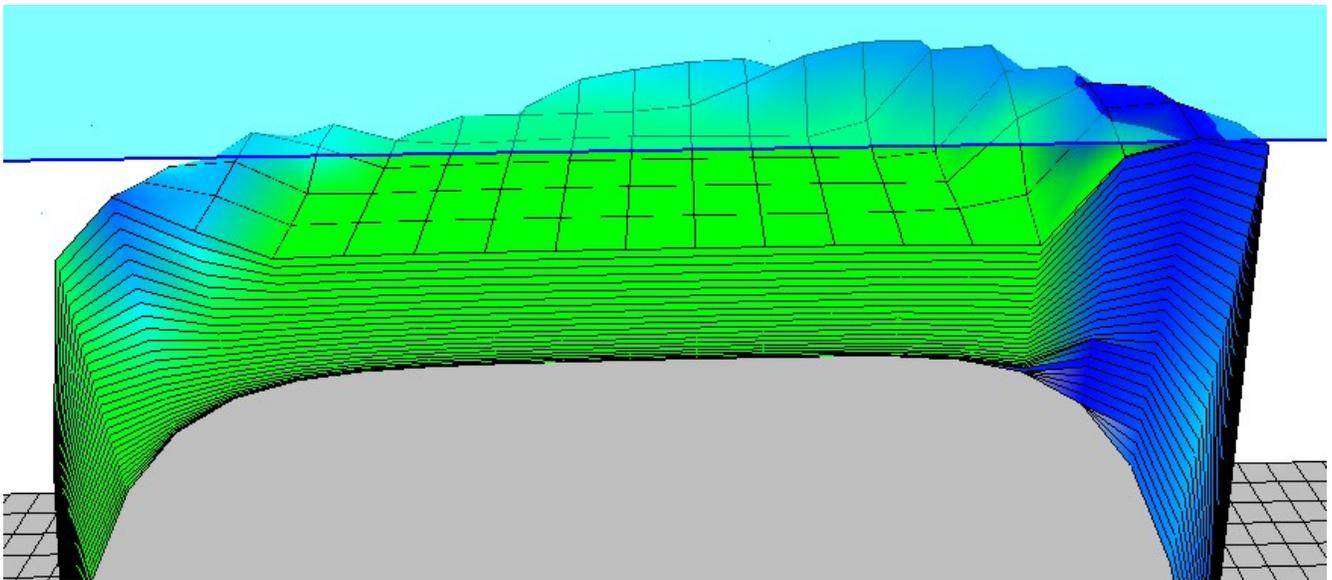

*Figure 31: Cross section through atoll model shown in previous figure.*

## Aggradation and deep-sea sedimentation

In order to provide for sediment input over a large area, or along the edge of a model, SedSimple provides a mechanism for the user to specify sediment input through maps and time curves. This can be used to model deep sea sedimentation between turbidite flows, volcanic ash deposition, or any mechanism for input of sediment from outside the area. Channelized flow also allows to input sediment, but this should be used only when a channel (river or turbidite) enters the model area at an edge.

The sediment supply can be specified as a rate that is areally uniform throughout the model, or varies with depth (specified as a depth dependent curve), or depends on the location (specified as a map).

Depth-dependency may be desirable for deep-sea sedimentation when there is assumed to be little or no production in shallow areas (for example foraminifera production). In that case, the production curve may look similar to the following:

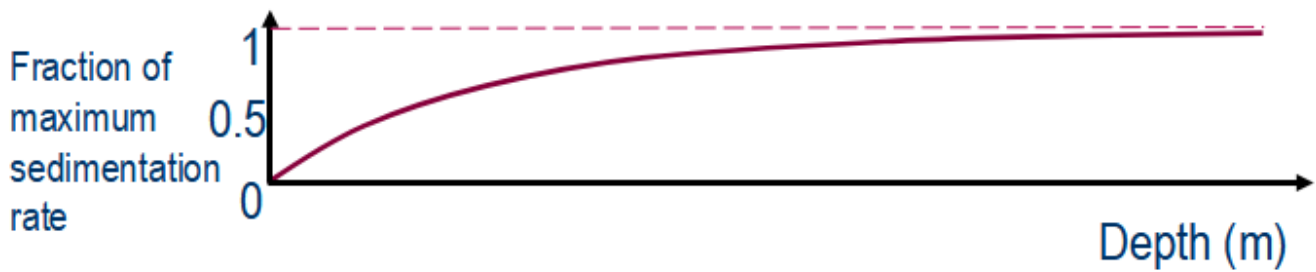

*Figure 32: Example of depth-dependent sedimentation rate.*

### Examples of aggradation

The following figures show example of uniform aggradation (single rate with no maps or curves) in combination with models presented previously in this paper.

The next figure shows the same fluvial model as Figure 22, but adding aggradation of fine sand and silt. The model is shown at an intermediate stage (75 ka of a total of 150 ka) to show how the river tends to form meanders, which does not occur in the original model)

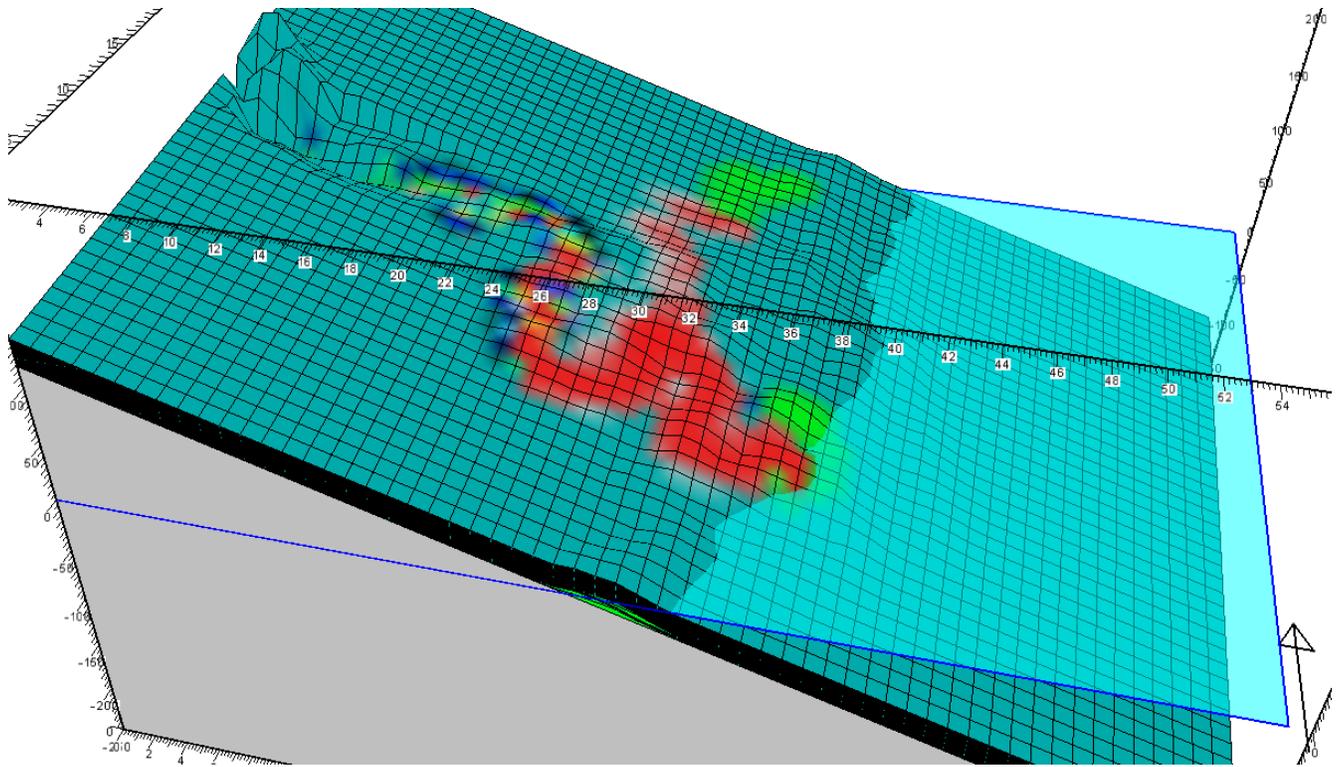

*Figure 33: Same model as previous figure, but with aggrdation. Intermediate result at 75 ka to show evolution of channel.*

A longitudinal cross section through the final model (150 ka) is shown below:

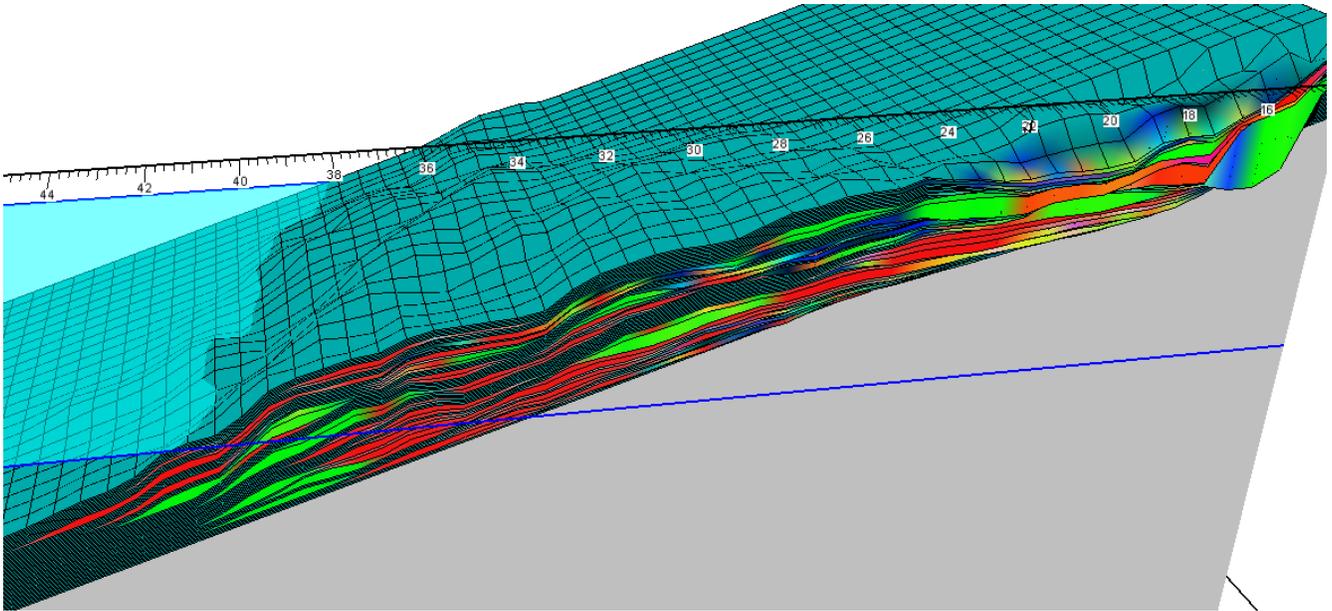

*Figure 34: Longitudinal section through fluvio-deltaic model with aggradation. Final result at 150 ka. Runtime = 1.05 seconds.*

The next figure shows the turbidite model of Figure 23, with the addition of fine sand and silt:

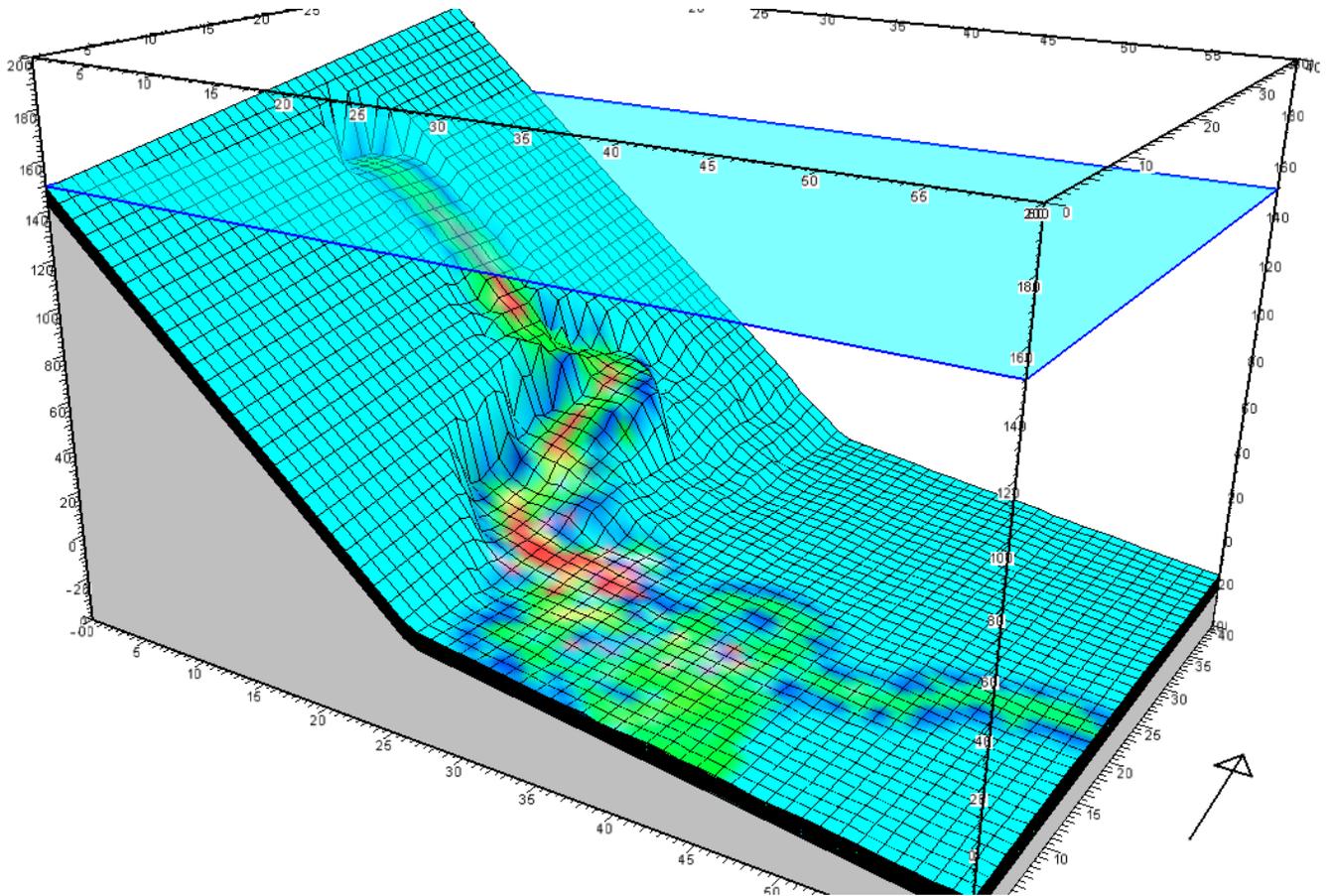
*Figure 35: Example of turbidite system with aggradation (areally uniform sediment input). Runtime = 0.60 seconds.*

A transversal section through the model shows a much more parallel architecture than the original model. This effect is probably exaggerated as in natural systems there would not be a large amount of sand and silt simple settling uniformly in the area, but the example serves to show how the stratigraphy may change by addition of aggradation.

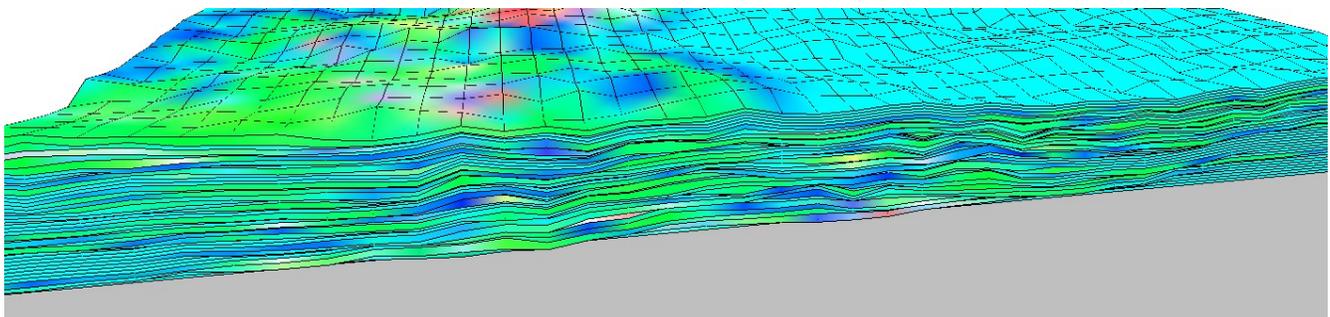
*Figure 36: Transversal section through turbidite deposits shown in previous figure.*

# Tectonics

Syntectonic sedimentation modeling can give rise to interesting stratigraphic features (Carmona et al., 2009, Carmona et al. 2016). SedSimple, is capable of modeling only vertical tectonic movement, such as vertical faults, gentle folds, or the vertical effect of salt domes on bathymetry to reroute turbidity currents.

However, it is important to notice that even in compressional margins in which inverse faults and complex faults are common, vertical tectonics may be sufficient to model at least some parts of a sequence. For example. In the follwing situation, the upper part of the model can be correctly modeled with vertical movement only:

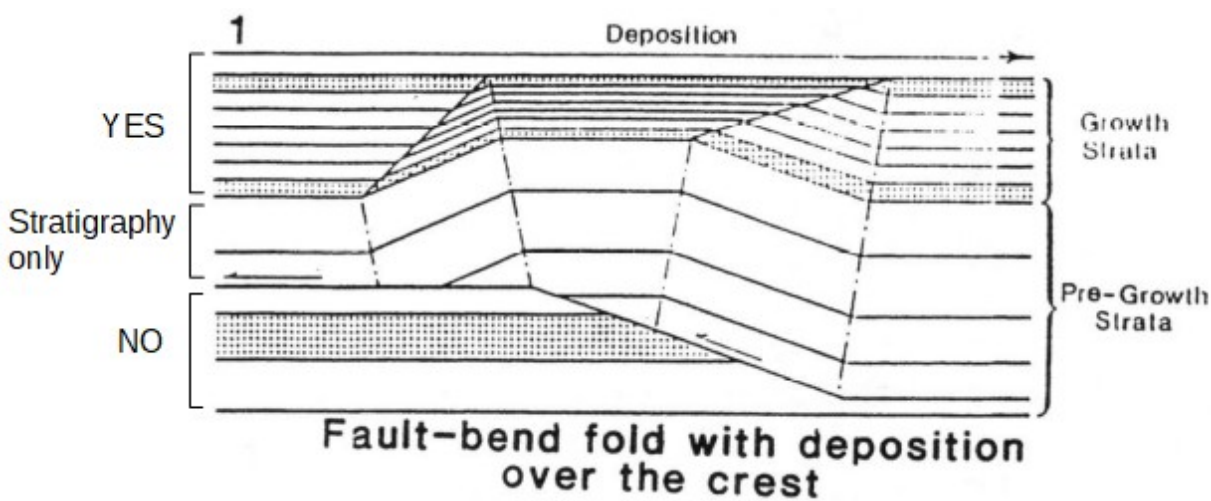

*Figure 37: Syntectonic deposition with vertical tectonics can still be accurate in certain portions of a sequence affected by low-angle faulting.*

In these cases, caution must be exercised to infer results only in the portion of the sequence that has been deposited while the tectonic regime at the surface was predominantly vertical.

In SedSimple, vertical tectonics are specified by means of maps (indicating vertical movement as a function of location) and time curves (indicating vertical movement as a function of time). Any number of maps and curves may be superimposed, thus providing unlimited capabilities to model any tectonic movement (though only vertical.

*Example of tectonic model*

The following figure shows a model of diffusion with two sea-level cycles with a vertical fault through the center of the model:

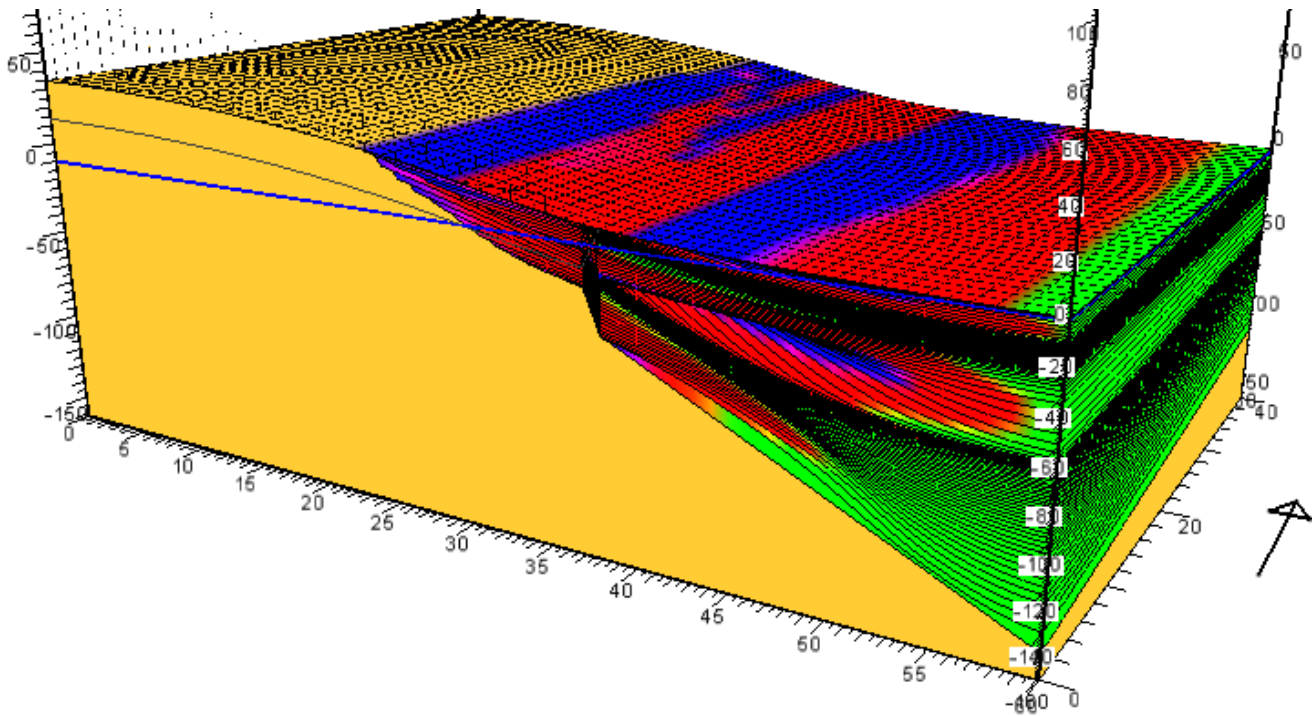
*Figure 38: Example of diffusion model with 2 sea-level cycles and a vertical fault through the center of the area.*

The next figure shows a turbidite system in which a mound (presumably caused by an underlying salt dome) rises during modeling. The left side shows the system without tectonic movement, while the right side shows the system with identical parameters except with a rising mound.

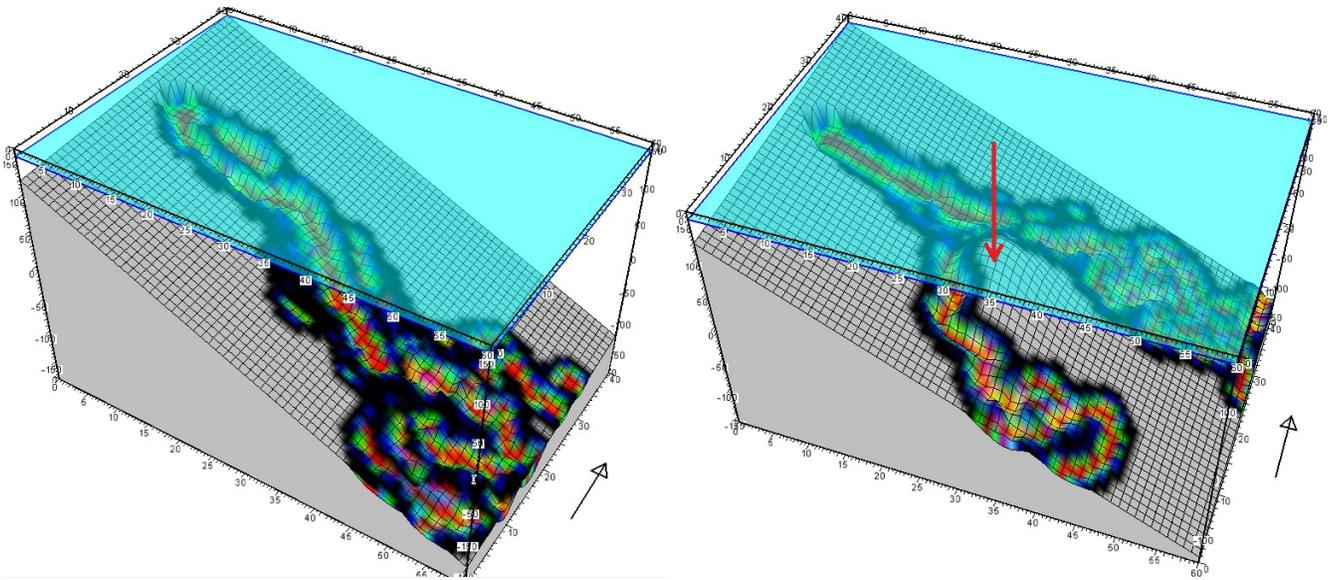

*Figure 39: Example of turbidite model. Left: no tectonics. Right: with a growing mound simulating a salt dome (red arrow) that causes channels to diverge.*

## Processes not yet modeled by SedSimple

SedSimple models many, but obviously not all, sedimentary and tectonic processes. Due to its flexibility it is likely that most remaining geologic process can be relatively easily implemented by users, or may be provided in the future as part of the standard release.

Among the processes not yet modeled are the following:

- Aeolian
- Glacial
- Evaporitic
- Wave-induced currents
- Carbonate drowning
- Tidal
- Deep sea currents, including contourites
- Mass transport (slides, slumps)
- Flocculation
- Compaction
- Isostatic compensation and lithospheric flexure
- Non-vertical tectonic deformation and complex faults

However, there is no fundamental reason why these processes could not be implemented. The software is well prepared to make it relatively simple to develop additional modules for these and other processes. It is likely that all of these processes will be incorporated in future releases.

# Uncertainty and data honoring

One of the drawbacks of geologic forward models is that they do not perform well at honoring data. Although it would be an easy exercise to modify a forward model to honor data, predictive ability away from data is lost in the process. Several methods have been explored to combine stratigraphic forward modeling with other methods to obtain better matches with data. SedSimple does not incorporate any of these methods, but its speed and flexibility make it suitable as a tool in testing new technologies (Slingerland, 1990, Tetzlaff, 1990).

These methods are discussed here because they all require large amounts of synthetic data (which SedSimple could potentially provide) in order to function properly.

Some of these methods are:

- Exhaustive search
- Optimization
- Combination with geostatistics
- Controlled chaos
- Artificial intelligence and machine learning

## Exhaustive search

Exhaustive search consists of generating a very large number of models and searching them to extract those that match observations. The family of extracted models can then be used to make statistical predictions about areas away from data. The following figure illustrates this principle:

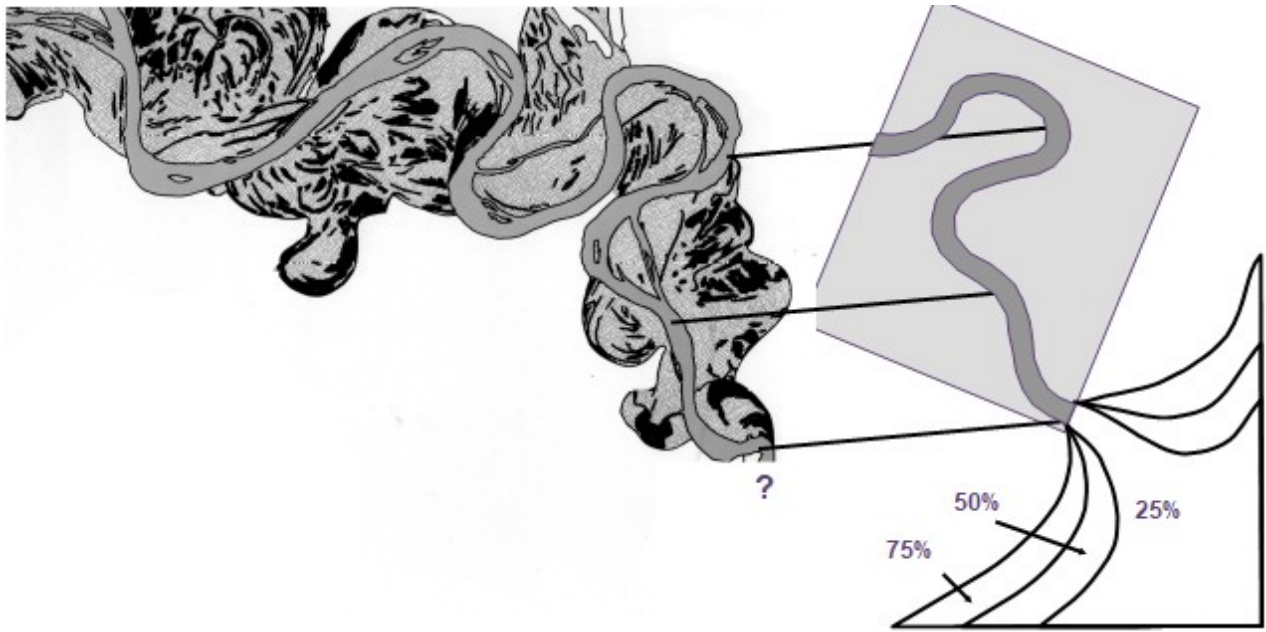

*Figure 40: Channel probability away from data by exhaustive search.*

*Optimization*

Optimization consist of varying the input parameters gradually in order to approach results that match data. This is done by means of an objective function, which is a function of the input parameters that provides a measure of the departure or error between the model and the data to honor. Finding the minimum (or minima) of this function should lead to a result that matches data

Optimization techniques have been used successfully in a number of modeling techniques. However, when the system is chaotic, they do not work well, because the objective function may be discontinuous at every point.

Various schematic objective functions (of a single variable for simplicitly) are shown in the next figure. It shows that in some cases (such as in chaotic systems) the minimum may not be discernible. However, in such cases it may still be possible to use the method successfully by relaxing the definition of error.

For example instead of attempting to match every well log at every level, we may just attempt to match net-to-gross, or a trend in the grain size over a large area, or the thickness of main sequences. In such cases the model may not be chaotic, and optimization methods may work.

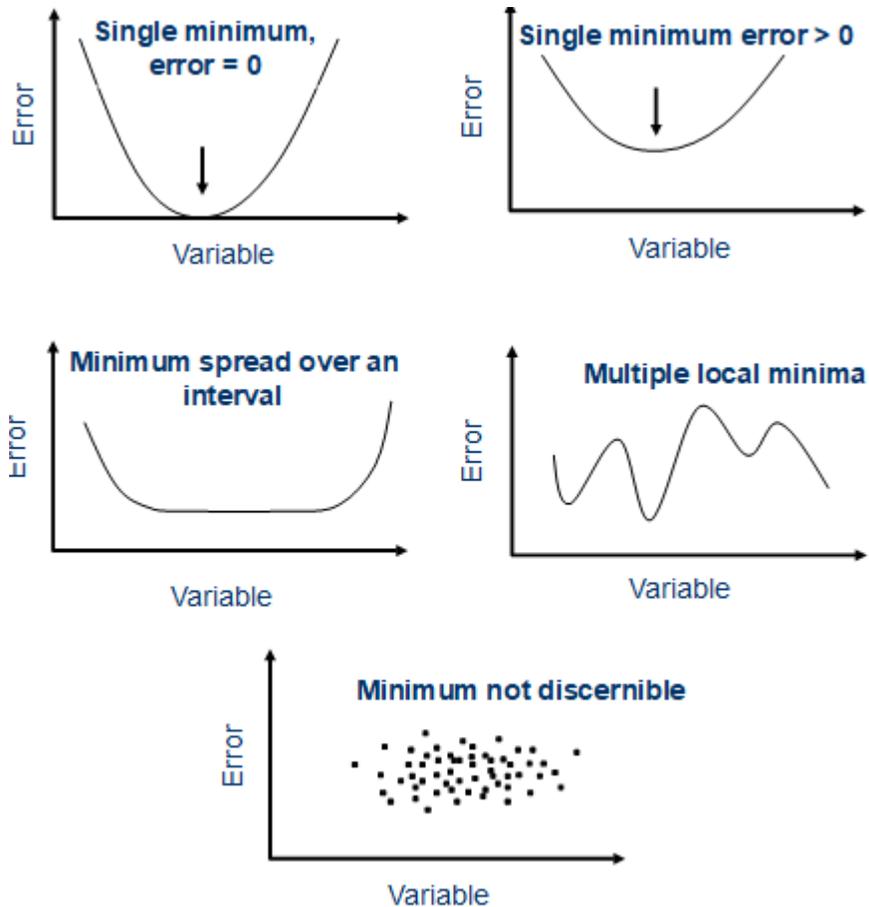
*Figure 41: Several possible objective functions.*

## Combination with geostatistics

Geostatistical methods are often presented as both an alternative and a complement to stratigraphic forward modeling methods (Doliguez et al., 1999). The following table briefly summarizes the main advantages and disadvantages of each set of methods:

| SFM Models | Geostatistical Models |
| --- | --- |
| Predict distribution of lithologies, even below seismic resolution | Can honor data in detail, such as exact petrophysical properties |
| Predict likely stratigraphy beyond data | Less CPU intensive |
| Integrate seismic, well-log, outcrop, and other geologic data | Do not depend on knowledge of past geologic conditions |
| Incorporates true geologic constraints | Simple forward workflow |

Another p[[possibility is to use geostatistics in combination with geologic forward modeling. This can take two forms, which can also be used together:

1. Extract variability (such as variograms or training images) from geologic model realizations and use them to obtain geostatistical realizations that honor data.

2. Use geologic model realizations (perhaps filtered through a low-pass filter) to obtain general trends of spatial variability. These trends are then used in a geostatistical model as "probability" cubes to better drive the simulation.

These methods have been used with some success in recent years. Multipoint geostatistics appears to be the geostatistical method of choice in these cases, as two-point (variogram) based geostatistics does not appear to capture the richness of variability produced by a geologic forward model. The latter point is illustrated by the following figures. The first figure is a realization of the "river" program (described later in this chapter) with aggradation, to yield a 3D mode and in particular this cross section. The next figure is a two-point unconditional geostatistical realization with the same variogram.

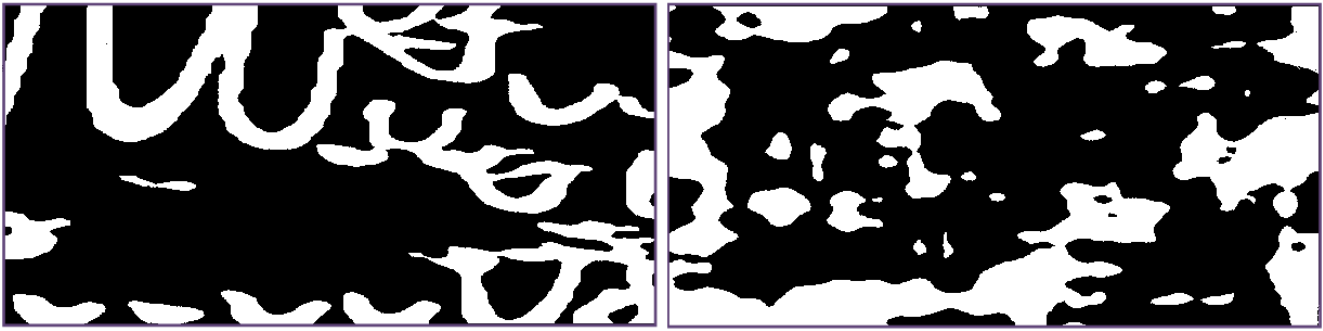

*Figure 42:    Left: Cross section through fluvial deposits modeled by a stratigraphic forward modeling program. Right: Unconditional variogram-based realization of the figure on the left.*

White represents channel (sand) and black intra channel (shale). These results show that two-point geostatistics may not be adequate to use in combination with geologic modeling, but multipoint geostatistics or other advanced methods may be (Fournillon et al., 2020).

*Controlled chaos*

One of the most interesting (but perhaps not fully realizable) possible developments in the quest to honor data with geologic models (particularly those that are chaotic) is "controlled chaos" (Tetzlaff, 1990).

Certain chaotic systems will turn periodic when forced by even a small periodic signal. This can be verified by the "river" program (described below). Once the system is periodic, small changes in the input should produce only small changes in the periodic output. Thus it may be possible to find a combination of parameters that produces a desired result in a purely chaotic model. However, this has not yet been achieved and may be difficult or impossible with very complex models.

The following figure shows the evolution of a realization of the "river" program with no perturbation.

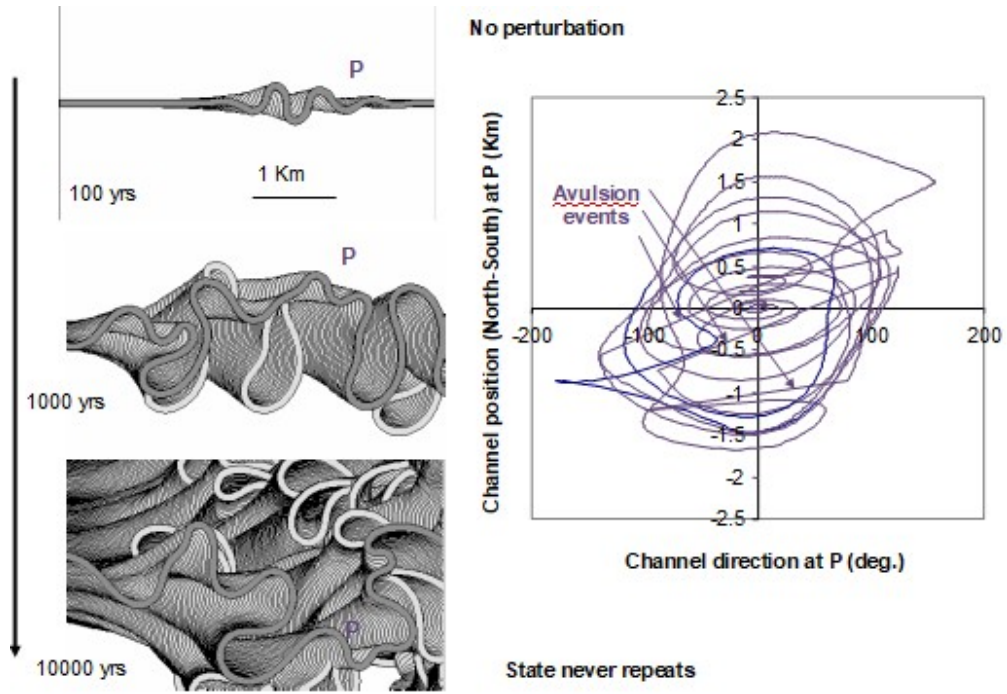

*Figure 43: Evolution of a fluvial channel exhibiting chaos.*

The next figure shows the evolution of the system when introducing a periodic perturbation. The system is no longer chaotic and may be changed gradually to "pass" through a deired point.

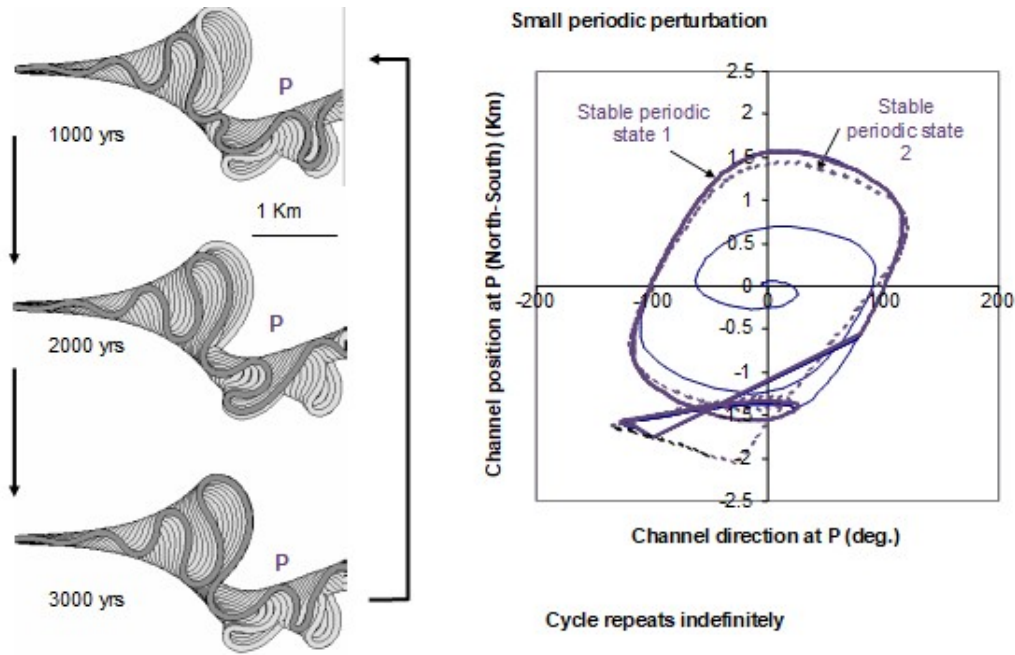

*Figure 44: Evolution of the fluvial channel shown in previous figure after introducing a low-amplitude forcing function, causing the model to become cyclic.*

Once the system is forced to behave periodically instead of chaotically, presumably optimization methods could be used to make the channel pass through certain data points.

*Artificial intelligence*

Artificial intelligence (AI) and related techniques (machine learning, neural networks) have been applied successfully in geology in the prediction and extrapolation from sparse data (Koroteev and Tekic, 2021). AI can detect and apply relationships between data without requiring the explicit intervention of the user for each specific case. Bayesian tomographic imaging and neural networks have been used in combination with stratigraphic forward modeling techniques to obtain geologically reasonable scenarios from travel time data (Bloem et al., 2022).

The field is vast and growing rapidly. No attempt will be made here to predict what role it will play in the interpretation of geologic systems, but there is no doubt that there are numerous applications. In the context of stratigraphic forward modeling,

AI might not become an "assistance" to stratigraphic forward modeling methods in order to honor hard and soft data. Rather, it may be the other way around: stratigraphic forward modeling may become an auxiliary to AI methods in generating large numbers of realizations that are base on physical and geological principles in the form of complete and detailed three-dimensional models, thereby bringing physical cause and effect relationships into the realm of AI.

SedSimple has been developed primarily with this goal in mind, and we hope that it will contribute to its fulfillment.

# Conclusions

1. The SedSimple package fulfills the need for a fast and flexible stratigraphic forward modeling package that can be tailored to specific needs, with full access to the source code.

2. Although the algorithms are relatively simple, the results are often comparable to those of major packages.

3. SedSimple appears to be adequate for fast experimentation with methods that require a large number of outputs.

4. For the user of conventional geologic modeling SedSimple can be used without the need to understand or even look at the code, or by a researcher or programmer interested in incorporating new algorithms to work in conjuction with others already implemented.

# Acknowledgements

I thank Westchase Software Corporation for allowing the publication of this paper. I am also indebted to Prof. Andrew Curtis and Hugo Bloem of the University of Edinburgh for applying SedSimple to their research on machine learning and suggesting valuable improvements.

# References

Bridge, J., 1975, "Computer simulation of sedimentation in meandering streams", Sedimentology, v. 22, pp. 3-43.

Bloem, H., A. Curtis, and D. Tetzlaff, 2022, "Introducing conceptual geological information into Bayesian tomographic imaging", http://www.arxiv.org, arXiv:2210.07892, 18pp.

Burns, P., and E. Meiburg, 2015, "Sediment-laden fresh water above salt water: nonlinear simulations", J. Fluid Mech. (2015), vol. 762, pp. 156195., Cambridge University Press 2014 doi: 10.1017/jfm 2014.645.

Carmona, A., R. Clavera-Gispert, O. Gratacós, S. Hardy, 2009, "Modelling Syntectonic Sedimentation: Combining a Discrete Element Model of Tectonic Deformation and a Process-based Sedimentary Model in 3D", Math Geosci (2010) 42: 519–534, DOI 10.1007/s11004-010-9293-6.

Carmona A, O. Gratacós, R. Clavera-Gispert, J. Muñoz, S. Hardy, 2016 "Numerical modelling of syntectonic subaqueous sedimentation: The effect of normal faulting and a relay ramp on sediment dispersal", Tectonophysics 684 (2016) 100–118.

Doliguez, B., D. Granjeon, P. Joseph, R. Eschard, and H. Beucher, 1999, How can stratigraphic modeling help constrain geostatistical reservoir simulations? *in* [2], p. 239-244.

Duan, T., T. Cross, and M. Lessenger, 2000, 3-D carbonate stratigraphic model based on energy and sediment flux. Poster Presentation - 2000 AAPG Annual Meeting.

Ferguson, R., 1984. "Kinematic Model of Meander Migration", *in*: Elliot, C. M., 1984 (ed.), River Meandering, Proceedings of the 1983 Conference on Rivers, New Orleans, American Society of Civil Engineers, 942-951.

Fournillon, S. Daher, E. Marfisi, N. Hawie, M. Callies, 2021, "Forward stratigraphic modelling at reservoir scale, statistical comparison with common geostatistical methods and fractal behaviour", EAGE, 82$^{nd}$ Annual Conference and Exhibition.

Fu, C., X. Yu, Y. He, J. Liang, Z. Kuang, 2019, "Different initial density turbidite sediments with coarse grain injection and their corresponding flow pattern: Additional insights from numerical simulation in a study case of South China Sea", Petroleum Research 4 (2019) 19e28.

Harbaugh, J. W., and Bonham-Carter, G., 1981, Computer simulation in geology, Krieger Publishing Co., Malabar, Florida, reprint edition 1981.

Heerema, C., P. Tallinga, M. Cartignya, C. Paullb, L. Bailey, S. Simmons, D. Parsons, M. Clare, R. Gwiazda, E. Lundsten, K. Anderson, K. Maier, J. Xu, E. Sumnerc, K. Rosenbergerf, J. Gales, M. McGann, L. Carter, E. Pope, 2020, "What determines the downstream evolution of turbidity currents?", Earth and Planetary Science Letters 532 (2020) 116023.

Hill, J., R. Wood, A. Curtis, and D. Tetzlaff, 2012, Preservation of forcing signals in shallow water carbonate sediments, Sedimentary Geology, vol. 275-276, pp 79-92.

Jeschke, S., and C. Wojtan, 2015, "Water Wave Animation via Wavefront Parameter Interpolation", ACM Transactions on Graphics, Vol. 34, No. 3, Article , Publication date: April 2015.

Kaufman, P., J. Grotzinger, and D. McCormick, 1991, "Depth-dependent Diffusion Algorithm for Simulation of Sedimentation in Shallow Marine Depositional Systems", in *Sedimentary Modeling: Computer Simulations and Methods for Improved Parameter Definition*, Kansas Geological Survey Bulletin, 233, 524 pp.

Koroteev, D., and Z. Tekic, 2021, "Artificial intelligence in oil and gas upstream: Trends, challenges, and scenarios for the future", Energy and AI, vol. 3, March 2021, 100041

Liu, Y., H. Chen, J. Wang, S. Yang, and A. Chen, 2020, "Numerical simulation for the effects of waves and grain size on deltaic processes and morphologies", Open Geosciences 2020; 12: 1286–1301.

Simon, B., C. Robin, D. Rouby, J. Braun, F. Guillocheau, 2022, "Estimating sediment transport diffusion coefficients from reconstructed rifted margin architecture: measurements in the Ogooué and Zambezi deltas", *Basin Research.* 2022;34:2064–2084.

Meiburg, E., S. Radhakrishnan, M. Nasr-Azadani, 2015, "Modeling Gravity and Turbidity Currents: Computational Approaches and Challenges", Applied Mechanics Reviews, ASME July 2015, Vol. 67 / 040802-1.

Nasr-Azadani, M. and E. Meiburg, 2014, "Turbidity currents interacting with three-dimensional sea-floor topography", J. Fluid Mech. (2014), vol. 745, pp. 409443, Cambridge University Press 2014

doi:10.1017/jfm.2014.47.

Parkinson, S., 2014, "Advances in computational modelling of turbidity currents using the finite-element method", PhD Thesis, Imperial College London Department of Earth Sciences & Engineering, 159 pp.

Shimizu, H., T. Koyaguchi and Y. Suzuki, 2017, "A numerical shallow-water model for gravity currents for a wide range of density differences", *Progress in Earth and Planetary Science* (2017) 4:8 DOI 10.1186/s40645-017-0120-2.

Slingerland, R., 1990. Predictability and chaos in quantitative dynamic stratigraphy. In: T. A. Cross (ed.): Quntitative dynamic stratigraphy. Prentice-Hall, Englewood Cliffs, New Jersey, 625 pp.

Sun T., P. Meakin, and T. Jossang, 1996. A Simulation Model of Meandering Rivers. Water Resources Research, **v. 32**, 2937-2954.

Tetzlaff, D. M., and Harbaugh, J. W., 1989, *Simulating Clastic Sedimentation*, van Nostrand Rheinold Series on Computer Methods in Geosciences, New York, 202 pp.

Tetzlaff, D. M., 1990, "Limits to the Predictive Ability of Dynamic Models that Simulate Clastic Sedimentation", *in*: Cross, T. (ed.), *Quantitative dynamic Stratigraphy"*, Prentice Hall, New Jersey, 625 pp.

## Software reference

This paper describes the program SedSimple, revision 1.03. Later versions will preserve the features described here but will likely add more processes. The program is licensed under the GNU General Public License and is available for download at:

> https://www.wsoftc.com

Both the source code as well an executable for Windows are available. It should be easy to build the program for other operating systems.

The companion graphics program (Sirius) is available for download under a no-cost proprietary license and is provided only as an executable for Windows.

For doubts, suggestions and questions, please use the "Contact us" form available at that web site.